\documentclass[11pt]{article}
\usepackage{a4wide}
\usepackage{mathrsfs}
\usepackage{latexsym,bm}
\usepackage{graphicx}
\usepackage{indentfirst}
\usepackage{slashed}
\usepackage{amsmath}
\usepackage{amssymb}
\usepackage[usenames,dvipsnames]{color}
\usepackage{hyperref}
\usepackage{epsfig}
\usepackage[titletoc]{appendix}
\usepackage{multirow}%
\usepackage{rotating}
\usepackage[square,numbers,sort&compress]{natbib}
\usepackage{tikz,tikz-feynman,graphbox,subfigure}
\usepackage{ulem} 
\usepackage[top=2.9cm,bottom=2.5cm,left=2.8cm,right=3cm]{geometry}
\usepackage{tabularx}

\newcommand{\email}[1]{\footnote{{\em } \texttt{#1}}}

\newcommand{\bma}{\left(\begin{matrix}}
\newcommand{\ema}{\end{matrix}\right)}

\newcommand{\bra}{\langle}
\newcommand{\ket}{\rangle}
\newcommand{\xpt}{{\chi}{\rm PT}}

\begin{document}

\thispagestyle{empty}
\title{
\Large \bf Axion-pion scattering at finite temperature in chiral perturbation theory and its influence in axion thermalization}
\author{\small Jin-Bao Wang$^a$,\,  Zhi-Hui Guo$^{b}$\email{zhguo@hebtu.edu.cn}, \, Hai-Qing Zhou$^{a}$\email{zhouhq@seu.edu.cn} \\[0.5em]
{ \small\it ${}^a$ School of Physics, Southeast University, Nanjing 211189, China }
\\[0.2em] 
{ \small\it ${}^b$ Department of Physics and Hebei Key Laboratory of Photophysics Research and Application, } \\ 
{\small\it Hebei Normal University,  Shijiazhuang 050024, China}
}
\date{}

%

\maketitle
\begin{abstract}
Axion-pion scattering amplitudes at finite temperatures are calculated within chiral perturbation theory up to the one loop level. Unitarization procedure is implemented to these amplitudes in order to extend the applicable range of energy and temperature. The influence of the thermal axion-pion scattering amplitudes on the $a\pi\to\pi\pi$ cross sections and the axion thermalization rate is investigated, with the emphasis on the comparison with the zero-temperature-amplitude case. A brief discussion on the cosmological implication of the axion thermalization rate, that is calculated by using the $a\pi\to\pi\pi$ amplitudes at finite temperatures, is also given. The thermal corrections to the axion-pion scattering amplitudes can cause around a 10\% shift of the determination of the axion decay constant $f_a$ and its mass $m_a$, comparing with the results by using the $a\pi\to\pi\pi$ amplitudes at zero temperature.  
\end{abstract}

\section{Introduction}

The conjectural particle axion constitutes a dynamical research subject in many areas of physics, including particle physics, cosmology, astronomy, etc~\cite{Kim:2008hd,Graham:2015ouw,Irastorza:2018dyq,DiLuzio:2020wdo,Choi:2020rgn,Sikivie:2020zpn}. The primary aim of the QCD axion is to solve the strong CP problem, which can be elegantly addressed within the Peccei-Quinn (PQ) mechanism~\cite{Peccei:1977hh,Weinberg:1977ma,Wilczek:1977pj}. The axion is conjectured to be a Nambu-Goldstone boson that arises from the spontaneous breaking of a global $U(1)_{PQ}$ symmetry at some high energy scale $f_a$, also known as the axion decay constant. The key idea of the PQ mechanism is to promote the anomalous term $\theta G\tilde{G}$ in QCD, being $G$ and $\tilde{G}$ the gluon filed tensor and its dual, by the dynamical axion field $a$ as $\frac{a}{f_a}G\tilde{G}$, which softly breaks the $U(1)_{PQ}$ symmetry and makes the axion actually a pesudo-Nambu-Goldstone boson (pNGB). In fact the anomalous $\frac{a}{f_a}G\tilde{G}$ term is commonly regarded as the model independent part in the construction of various axion models. Being the essential QCD nature, axion inevitably interacts with the hadrons and intensive investigations are carried out to find evidence for such interactions~\cite{DiLuzio:2020wdo}.

One attractive way to trace the axion imprint is to study the axion production in the thermal medium of the early Universe. Below the critical temperature ($T_c\sim 150$~MeV) of QCD phase transition, the quarks and gluons are confined inside hadrons and consequently one needs the axion-hadron interactions as inputs to reliably calculate the axion production in the thermal bath. As the lightest QCD hadron, the pions are expected to provide the most important sources for the axion production below $T_c$. Indeed, continuous efforts have been made to refine the axion-pion interactions in recent years, in order to improve the cosmological constraints on the axion properties. In a pioneer work~\cite{Chang:1993gm}, the axion-pion scattering amplitude is calculated by taking the leading-order (LO) chiral perturbation theory ($\xpt$). The LO $\xpt$ axion-pion amplitude is adopted in many sequential works to constrain the axion decay constant $f_a$ or its mass $m_a$ \cite{Hannestad:2005df,Hannestad:2007dd,Melchiorri:2007cd,Giusarma:2014zza,DEramo:2021psx}. Recently, the calculation of the axion-pion scattering amplitude is pursued up to the next-to-leading order (NLO) in $\xpt$, and an important finding is that the perturbative $\xpt$ results become unreliable for the temperature $T>62$~MeV, due to the loss of convergence of the chiral expansion for the axion thermalization rate~\cite{DiLuzio:2021vjd}. Later on, different strategies are proposed to extend the applicable region of the $\xpt$ amplitudes in Refs.~\cite{Notari:2022ffe,DiLuzio:2022gsc}. The inverse amplitude method (IAM) is employed in Ref.~\cite{DiLuzio:2022gsc} to unitarize the NLO axion-pion amplitude. While, in Ref.~\cite{Notari:2022ffe} it is assumed that the axion-pion amplitudes can be approximated by scaling the pion-pion amplitudes with the axion-pion mixing strength factor. It is pointed out that in all the previous works the axion thermalization rates are calculated by taking the axion-pion scattering amplitudes at zero temperature, instead of the thermal ones, as an approximation. In this work, we fill this gap by performing the $a\pi \to \pi\pi$ scattering amplitudes at finite temperatures in $\xpt$ up to the one-loop level. Then we briefly study the influence of the thermal $a\pi\to\pi\pi$ amplitudes on the axion thermalization rates and the determinations of the axion parameters by confronting the extra effective number of relativistic thermal degrees of freedom $\Delta N_{\rm eff}$.

This article is organized as follows. In Sec.~\ref{sec.lag}, we elaborate the relevant axion chiral Lagrangian and the calculation of $a\pi\to\pi\pi$ scattering amplitudes at finite temperatures up to the one-loop level. A survey of thermal corrections to the scattering amplitudes will be also given in this section, with the emphasis on comparing with the zero-temperature amplitudes used in literature. In Sec.~\ref{sec.cosmo}, a brief discussion about the cosmological constraints on the axion parameters relying on the averaged axion thermalization rates calculated from the thermal $a\pi\to\pi\pi$ scattering amplitudes will be presented. We give a short summary and conclusions in Sec.~\ref{sec.concl}. Technical details about the calculations of thermal loops and phase space integrals are given in Appendices.

\section{Axion-pion scattering amplitudes at finite temperatures}\label{sec.lag}

\subsection{Relevant axion $\xpt$ Lagrangians}

In this work we stick to the minimal QCD axion Lagrangian 
\begin{equation}
\mathcal{L}_{a\text{QCD}}= \mathcal{L}_{\text{QCD}} +  \frac{1}{2}\partial_{\mu}a\partial^{\mu}a
+\frac{a}{f_a}\frac{g_s^2}{32\pi^2}G_{\mu\nu}^c\widetilde{G}^{\mu\nu}_c\,,
\end{equation}
with the dual of the gluon tensor $\widetilde{G}^{\mu\nu}_c\equiv\varepsilon^{\mu\nu\rho\sigma}G_{\rho\sigma}^c/2$. Various extensions of the axion interactions with quarks, leptons and photons can be found in many recent reviews~\cite{Kim:2008hd,Graham:2015ouw,Irastorza:2018dyq,DiLuzio:2020wdo,Choi:2020rgn,Sikivie:2020zpn}.
By performing the quark field transformation $q\to e^{i\frac{a}{2f_a}Q_a \gamma_5 } q$, being $Q_a$ a $2\times 2$ hermite matrix in quark flavor space with $\mathrm{Tr}(Q_a)=1$, one can eliminate the anomalous term $aG\widetilde{G}$ and in the meantime also introduce additional axial-vector coupling 
\begin{equation}\label{eq.avqcurrent}
-\frac{\partial_\mu a}{2 f_a} \bar{q} \gamma^\mu \gamma_5 Q_a q \,,
\end{equation}
and a modification to the quark mass term as
\begin{equation}
 -\bar{q}\, e^{i\frac{a}{2f_a}Q_a\gamma_5} M_q e^{i\frac{a}{2f_a}Q_a\gamma_5}\, q \,,
\end{equation}
where $M_q$ corresponds to the two-flavor quark mass matrix $M_q=\text{diag}(m_u,\,m_d)$\,. 

Based on these ingredients, the axion $\xpt$ Lagrangian can be constructed order by order as the usual $\xpt$~\cite{Gasser:1983yg,Gasser:1984gg}
\begin{equation}
\mathcal{L}_{a\xpt}=\frac{1}{2}\partial_{\mu}a\partial^{\mu}a+\mathcal{L}_2+\mathcal{L}_4+\cdots\,.
\end{equation}
The LO axion $\xpt$ Lagrangian at $\mathcal{O}(p^2)$ in the $SU(2)$ reads
\begin{equation}
\mathcal{L}_2=\frac{F^2}{4}\langle \partial_{\mu}U\partial^{\mu}U^{\dagger}+\chi_aU^{\dagger}+U\chi_a^{\dagger}\rangle
+\frac{\partial_{\mu}a}{2f_a}J^{\mu}_A\big|_{\text{LO}}\,,\label{eq.L2}
\end{equation}
where $F$ corresponds to the pion decay constant at LO, $\chi_a=2B_0M_q(a)$, and the axion dressed quark mass matrix $M_q(a)$ is given by 
\begin{equation}\label{eq.defmqa}
M_q(a)=e^{-i\frac{a}{2f_a}Q_a} M_q e^{-i\frac{a}{2f_a}Q_a}\,. 
\end{equation}
The pion fields are collected in $U=e^{i \sqrt{2}\Phi/F}$ with 
\begin{align}
\Phi= \left( \begin{array}{cccc}
\frac{\pi^0}{\sqrt{2}} & \pi^+ \\
\pi^- & -\frac{\pi^0}{\sqrt{2}}
\end{array} \right) \,.
\end{align}
$J^{\mu}_A\big|_{\text{LO}}$ stands for the axial-vector current $-\bar{q} \gamma^\mu \gamma_5 Q_a q$ in Eq.~\eqref{eq.avqcurrent} at the hadron level from the LO $\xpt$
\begin{equation}\label{eq.amulo}
J^{\mu}_A\big|_{\text{LO}}=-i\frac{F^2}{2}\big\langle Q_a\left(\partial^{\mu}UU^{\dagger}+U^{\dagger}\partial^{\mu}U\right)\big\rangle\,.
\end{equation}
In literature sometimes different assignments for the chiral transformation behaviors of the chiral building blocks, especially the axion dressed quark mass terms, are assumed in the calculation. Here we follow the conventions of the seminal $\xpt$ works in Refs.~\cite{Gasser:1983yg,Gasser:1984gg} for the chiral transformation behaviors of the various building blocks: $\chi\equiv 2B_0(s+ip)\to \chi'= V_R \chi V_L^\dagger$ (so that the QCD quark mass term $-\bar{q}_{_R} (s+ip) q_{_L} + {\rm h.c.}$ is chiral invariant), and $U\to U'=V_R U V_L^\dagger$, the latter of which implies that the covariant derivative takes the form $D_\mu U= \partial_\mu U - i r_\mu U + i U l_\mu$, with the relations between the right/left-hand ($r_\mu/l_\mu$) and vector/axial-vector ($v_\mu/a_\mu$) external sources: $r_\mu=v_\mu+a_\mu$ and $l_\mu=v_\mu-a_\mu$. The sign conventions introduced in Eq.~\eqref{eq.defmqa} for the axion dressed quark mass and Eq.~\eqref{eq.amulo} for the axial-vector current are consistent with the just mentioned definitions of chiral transformation behaviors used in Refs.~\cite{Gasser:1983yg,Gasser:1984gg}. It is mentioned that there is a minus sign difference for the definition of hadronic axial-vector currents between ours in Eq.~\eqref{eq.amulo} and the one in Ref.~\cite{DiLuzio:2022gsc}.

Regarding the choice of $Q_a$, it is noticed in Ref.~\cite{Georgi:1986df} that by taking 
\begin{equation}\label{eq.qa}
Q_a= \frac{M_q^{-1}}{\bra M_q^{-1} \ket}\,,
\end{equation}
the mass mixing between axion and pion will be automatically eliminated at LO. Nevertheless, the last term in Eq.~\eqref{eq.L2} will lead to the kinematic mixing for axion and pion. To be definite in our calculation, we will take the form of $Q_a$ in Eq.~\eqref{eq.qa} throughout, although the physical quantities should remain the same with different choices of $Q_a$.  

The axion-pion scattering amplitudes receive the NLO contributions at $\mathcal{O}(p^4)$ both from the loops by taking the LO vertices in Eq.~\eqref{eq.L2} and the $\mathcal{O}(p^4)$ local chiral operators~\cite{Gasser:1983yg,Gasser:1987rb} 
\begin{equation}
\begin{aligned}
\mathcal{L}_{4}=
&\frac{l_3+l_4}{16}\langle\chi_aU^{\dagger}+U\chi_a^{\dagger}\rangle\langle\chi_aU^{\dagger}+U\chi_a^{\dagger}\rangle
+\frac{l_4}{8}\langle \partial_{\mu}U\partial^{\mu}U^{\dagger}\rangle\langle\chi_aU^{\dagger}+U\chi_a^{\dagger}\rangle
\\
&-\frac{l_7}{16}\langle\chi_aU^{\dagger}-U\chi_a^{\dagger}\rangle\langle\chi_aU^{\dagger}-U\chi_a^{\dagger}\rangle
+\frac{h_1-h_3-l_4}{16}\left[
\left(\langle\chi_aU^{\dagger}+U\chi_a^{\dagger}\rangle\right)^2
\right.
\\
&\left.
+\left(\langle\chi_aU^{\dagger}-U\chi_a^{\dagger}\rangle\right)^2
-2\langle\chi_aU^{\dagger}\chi_aU^{\dagger}+U\chi_a^{\dagger}U\chi_a^{\dagger}\rangle
\right]+\frac{\partial_{\mu}a}{2f_a}J^{\mu}_A\big|_{\text{NLO}}\,,\label{eq.lag4}
\end{aligned}
\end{equation}
where the NLO piece of the axial-vector current is 
\begin{align}\label{eq.lag4av}
J^{\mu}_A\big|_{\text{NLO}}=
&\,- il_1\langle Q_a \left\{\partial^{\mu}U,\,U^{\dagger}\right\}\rangle\langle \partial_{\nu}U\partial^{\nu}U^{\dagger}\rangle 
-i\frac{l_2}{2}\langle Q_a\left\{\partial_{\nu}U,\,U^{\dagger}\right\}\rangle
\langle \partial^{\mu}U\partial^{\nu}U^{\dagger} + \partial^{\nu}U\partial^{\mu}U^{\dagger}\rangle\notag
\\
&-i\frac{l_4}{4}\langle Q_a\left\{\partial^{\mu}U\,,U^{\dagger}\right\}\rangle
\langle\chi_aU^{\dagger}+U\chi_a^{\dagger}\rangle\,.
\end{align}
Notice that only the terms relevant to this work are kept in Eqs.~\eqref{eq.lag4} and \eqref{eq.lag4av}.

We use the imaginary time approach to include the finite-temperature effects~\cite{Kapusta:2006pm,Bellac:2011kqa,Laine:2016hma}. Within the framework of $\xpt$~\cite{Gasser:1986vb,Gerber:1988tt}, the thermal corrections only enter via the chiral loops, while the low energy constants (LECs) $l_i$ and $h_i$ accompanying the local operators in Eqs.~\eqref{eq.lag4} and \eqref{eq.lag4av} do not depend on the temperatures. This in turn implies that once the values of the unknown LECs are fixed at zero temperature one can make pure predictions at finite temperatures. 
In Appendix~\ref{appendix.A}, we give relevant formulas for the one-loop integrals at finite temperatures.

\subsection{Calculation of axion-pion scattering amplitudes at finite temperatures up to one loop}

For the calculation of the amplitudes, one needs to address the LO $a$-$\pi^0$ mixing first. After taking the specific form of $Q_a$ in Eq.~\eqref{eq.qa}, the $a$-$\pi^0$ mixing at LO is exclusively caused by the $\partial_\mu a J^{\mu}_A\big|_{\text{LO}}$ term in Eq.~\eqref{eq.L2}, which leads to $\delta_{a\pi}\partial_\mu a \partial^\mu \pi^0$ with
\begin{equation}\label{eq.11}
\delta_{a\pi}=\frac{\delta_I\,F}{2f_a}\,,
\qquad
\delta_I=\frac{m_d-m_u}{m_d+m_u}\,.
\end{equation}
By taking the field redefinition: $a \to a-\delta_{a\pi}\pi^0+\mathcal{O}(1/f_a^2)$, $\pi^0 \to \pi^0+\mathcal{O}(1/f_a^2)$\,, one can eliminate the mixing term of axion and $\pi^0$ at LO and in the meantime render the coefficients of their kinetic terms as $1/2$ at the level of $\mathcal{O}(1/f_a)$\,(see Appendix~\ref{appendix.B}). 
Nevertheless, the $a$-$\pi^0$ mixing will appear again at NLO in the chiral expansion, which leads to a non-diagonal two-point functions $G_{ij}$ with $i,j=a,\pi^0$\,. In Appendix~\ref{appendix.B}, we give a detailed discussion about the two-point functions up to NLO, along with the temperature dependence of the axion and pion masses. 
The terms of $\mathcal{O}(1/f_a^2)$ will be neglected in the calculation of the axion-pion scattering amplitudes throughout.

\begin{figure}[t]
	\centering
	\subfigure[]{
		\includegraphics[align=c,scale=0.7]{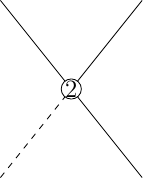}\label{fig.1a}
	}
	\subfigure[]{
		\includegraphics[align=c,scale=0.7]{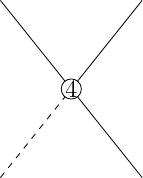}\label{fig.1b}
	}
	\subfigure[]{
		\includegraphics[align=c,scale=0.7]{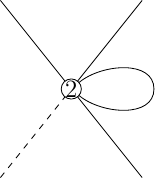}\label{fig.1c}
	}
	\subfigure[]{
		\includegraphics[align=c,scale=0.7]{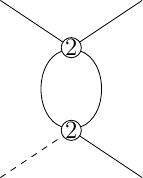}\label{fig.1d}
	}
	\subfigure[]{
		\includegraphics[align=c,scale=0.7]{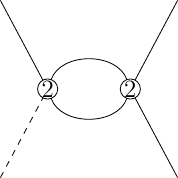}\label{fig.1e}
	}
	\subfigure[]{
		\includegraphics[align=c,scale=0.7]{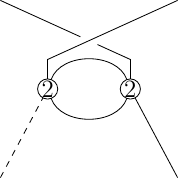}\label{fig.1f}
	}
	\subfigure[]{
		\includegraphics[align=c,scale=0.7]{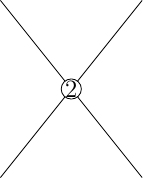}\label{fig.1g}
	}
	\caption{Feynman diagrams of the amputated four-point Green functions in Eqs.~\eqref{eq.ampMpm} and \eqref{eq.ampM00}. The dashed line stands for the axion field and the solid line corresponds to the pion field. The number $n$ for each vertex denotes its chiral order of $\mathcal{O}(p^{n})$. Diagrams \ref{fig.1a}-\ref{fig.1f} denote $A_{a\pi;\pi\pi}^{(2,4)}$, and diagram \ref{fig.1g} corresponds to $A_{\pi\pi;\pi\pi}^{(2)}$ which will be multiplied by the NLO $a$-$\pi^0$ mixing to give one type of contributions to the $a\pi\to\pi\pi$ scattering amplitudes at $\mathcal{O}(p^4)$, i.e. the last terms in Eqs.~\eqref{eq.ampMpm} and \eqref{eq.ampM00}.}
	\label{fig.FeynmanDiagrams}
\end{figure}

The axion-pion scattering amplitudes can be extracted from the four-point Green functions by using the Lehmann-Symanzik-Zimmermann (LSZ) reduction formula, as done in  Ref.~\cite{DiLuzio:2021vjd}. Retaining the contributions up to NLO, the $a\pi^0 \to \pi^+\pi^-$ and $a\pi^0 \to \pi^0\pi^0$ scattering amplitudes read
\begin{align}
&\mathcal{M}_{a\pi^0;\pi^+\pi^-}=\left(1+\frac{3}{2}\Sigma_{\pi\pi}^{(4)'}(m_{\pi}^2)\right)A_{a\pi^0;\pi^+\pi^-}^{(2)}+A_{a\pi^0;\pi^+\pi^-}^{(4)}-\frac{\Sigma_{a\pi^0}^{(4)}(0)}{m_{\pi}^2}A_{\pi^0\pi^0;\pi^+\pi^-}^{(2)}\,,\label{eq.ampMpm}
\\
&\mathcal{M}_{a\pi^0;\pi^0\pi^0}=\left(1+\frac{3}{2}\Sigma_{\pi\pi}^{(4)'}(m_{\pi}^2)\right)A_{a\pi^0;\pi^0\pi^0}^{(2)}+A_{a\pi^0;\pi^0\pi^0}^{(4)}-\frac{\Sigma_{a\pi^0}^{(4)}(0)}{m_{\pi}^2}A_{\pi^0\pi^0;\pi^0\pi^0}^{(2)}\,,\label{eq.ampM00}
\end{align}
and the other two charged channels $a\pi^+ \to \pi^+\pi^0$ and $a\pi^- \to \pi^0\pi^-$ are connected to $a\pi^0 \to \pi^+\pi^-$ by crossing symmetry.
The quantities of $A_{a\pi;\pi\pi}^{(2,4)}$ and $A_{\pi\pi;\pi\pi}^{(2)}$ in Eqs.~(\ref{eq.ampMpm}) and (\ref{eq.ampM00}) stand for the amputated four-point Green functions by taking the physical masses for the external states. $\Sigma_{\pi\pi}^{(4)}$ and $\Sigma_{a\pi}^{(4)}$ are the two-point one-particle-irreducible (1PI) amplitudes whose expressions can be found in Appendix~\ref{appendix.B}, and $\Sigma_{\pi\pi}^{(4)'}$ stands for the derivative of $\Sigma_{\pi\pi}^{(4)}$. The superscripts $(2)$ and $(4)$ denote the chiral orders. 
The corresponding Feynman diagrams for the amputated amplitudes $A_{a\pi;\pi\pi}^{(2,4)}$ and $A_{\pi\pi;\pi\pi}^{(2)}$ can be seen in Fig.~\ref{fig.FeynmanDiagrams}. The NLO amplitudes consist two parts: the chiral loops with the LO vertices~\eqref{eq.L2} and the pieces contributed by the $\mathcal{O}(p^4)$ local operators.

For the $a(p_1) \pi(p_2)\to\pi(p_3)\pi(p_4)$ scattering amplitude at finite temperatures, it can be decomposed into two parts: the zero temperature part and the thermal correction one, 
\begin{equation}
\mathcal{M}_{a\pi;\pi\pi}= \mathcal{M}_{a\pi;\pi\pi}^{(T=0)} + \Delta\mathcal{M}_{a\pi;\pi\pi}^{(T)}\,. 
\end{equation}
The amplitude $\mathcal{M}_{a\pi;\pi\pi}^{(T=0)}$ at zero temperature up to one loop has been computed and then plugged into the phase space integrals weighted by the Bose-Einstein (BE) distribution factors to estimate the axion thermal rates in Refs.~\cite{DiLuzio:2021vjd,DiLuzio:2022gsc}. As an improvement, we calculate the thermal correction part $\Delta\mathcal{M}_{a\pi;\pi\pi}^{(T)}$ up to one loop in this work, and its full expressions are given in Appendix~\ref{appendix.C}.
It is noted that at finite temperatures the commonly adopted Mandelstam variables
\begin{equation}
s=(p_1+p_2)^2\,, \quad t=(p_1-p_3)^2\,, \quad u=(p_1-p_4)^2\,, 
\end{equation}
are not enough to describe the thermal amplitudes $\Delta\mathcal{M}_{a\pi;\pi\pi}^{(T)}$, due to the loss of the Lorentz invariance at finite temperature $T$.
As shown later in Appendix~\ref{appendix.C} it is convenient to introduce the four momenta $p_s,\, p_t, \,p_u$, corresponding to $s,\,t,\,u$, respectively, to describe the thermal amplitudes~\cite{GomezNicola:2002tn}
\begin{equation}
p_s=p_1+p_2\,,
\quad
p_t=p_1-p_3\,,
\quad
p_u=p_1-p_4\,.
\end{equation}
In general the thermal amplitude can depend on the temporal and spatial components of the four momenta in a separate manner, as can be seen in Appendix~\ref{appendix.C}. 
Due to the rotation invariance, the number of independent kinematic variables in the finite-temperature amplitudes is five, and this number is also the effective dimension of the phase space integral in the evaluation of axion rate. In Appendix~\ref{appendix.D} we will illustrate in detail about the five specific kinematic variables to calculate the phase space integral.

Due to the momentum expansion feature of $\xpt$, its amplitudes are not expected to be convergent well at relatively high energy regions, especially where the resonances can appear. In Ref.~\cite{DiLuzio:2021vjd}, it is demonstrated that the axion thermalization rates calculated from the NLO $\xpt$ amplitudes are questioned to be valid above $T>62$~MeV. 
One efficient way to extend the applicable regions of energy and temperature for the perturbative $\xpt$ amplitudes is to restore the unitarity relations of the partial-wave (PW) amplitudes. We will use the inverse amplitude method (IAM)~\cite{Truong:1988zp,Dobado:1989qm,Dobado:1992ha,Oller:2020guq} to perform such unitarization, which will be done in the isospin ($I$) bases
\begin{align}
&\mathcal{M}_{a\pi^0;I=0}=-\frac{2}{\sqrt{3}}\mathcal{M}_{a\pi^0;\pi^+\pi^-}-\frac{1}{\sqrt{3}}\mathcal{M}_{a\pi^0;\pi^0\pi^0}\,,\label{eq.isoam00}
\\
&\mathcal{M}_{a\pi^0;I=2}=-\sqrt{\frac{2}{3}}\mathcal{M}_{a\pi^0;\pi^+\pi^-}+\sqrt{\frac{2}{3}}\mathcal{M}_{a\pi^0;\pi^0\pi^0}\,,\label{eq.isoam02}
\\
&\mathcal{M}_{a\pi^+;I=1}=-\sqrt{\frac{1}{2}}\mathcal{M}_{a\pi^+;\pi^+\pi^0}+\sqrt{\frac{1}{2}}\mathcal{M}_{a\pi^+;\pi^0\pi^+}\,,\label{eq.isoamp1}
\\
&\mathcal{M}_{a\pi^+;I=2}=-\sqrt{\frac{1}{2}}\mathcal{M}_{a\pi^+;\pi^+\pi^0}-\sqrt{\frac{1}{2}}\mathcal{M}_{a\pi^+;\pi^0\pi^+}\,,\label{eq.isoamp2}
\end{align}
where the conventions $|I,I_3\ket$ of the pions: $|\pi^+\ket=-|1,1\ket$, $|\pi^-\ket=|1,-1\ket$ and $|\pi^0\ket=|1,0\ket$ are used to obtain the above relations. Slight differences of the isotensor amplitudes in Eqs.~\eqref{eq.isoam02} and \eqref{eq.isoamp2} arise due to the isospin breaking effects in the axio-pion interactions.

At finite temperature $T$, we will work in the center of mass (CM) frame to perform the PW expansion 
\begin{equation}\label{eq.pwe}
\mathcal{M}_{a\pi,I}(E_{cm},\cos\theta)=\sum_{J}(2J+1)P_J(\cos\theta)\mathcal{M}_{a\pi;IJ}(E_{cm})\,,
\end{equation}
in order to proceed the unitarization procedure, with $P_J(\cos\theta)$ the Legendre polynomials. In the CM frame all the kinematic variables appearing in the thermal amplitudes can be related to the energy $E_{cm}$ and the scattering angle $\theta$ defined in that frame. The PW amplitudes with definite angular momentum $J$ can be obtained via Eq.~\eqref{eq.pwe} by performing the PW integrals
\begin{equation}
\mathcal{M}_{a\pi;IJ}(E_{cm})=\frac{1}{2}\int_{-1}^{+1}\mathrm{d}\cos\theta\,\mathcal{M}_{a\pi;I}(E_{cm},\cos\theta)P_J(\cos\theta)\,. 
\end{equation} 
We follow the IAM procedure to construct the unitarized PW $a\pi\to\pi\pi$ amplitude at finite temperature
\begin{equation}\label{eq.ampiam}
\mathcal{M}_{a\pi;IJ}^{\text{IAM}}=\frac{\left(\mathcal{M}_{a\pi;IJ}^{(2)}\right)^2}{\mathcal{M}_{a\pi;IJ}^{(2)}-\mathcal{M}_{a\pi;IJ}^{(4)}}\,,
\end{equation}
where $\mathcal{M}_{a\pi;IJ}^{(2)}$ and $\mathcal{M}_{a\pi;IJ}^{(4)}$ denote the $\xpt$ amplitudes at $\mathcal{O}(p^2)$ and $\mathcal{O}(p^4)$, respectively. 
The $s$-channel thermal unitariy relation of the IAM amplitude $\mathcal{M}_{a\pi;IJ}^{\text{IAM}}$ takes the form 
\begin{equation}
\text{Im}\mathcal{M}_{a\pi;IJ}(E_{cm})=\frac{\sigma_{\pi}(E_{cm}^2)}{32\pi}
\left[1+2n_B\bigg(\frac{E_{cm}}{2}\bigg)\right]\mathcal{M}_{\pi\pi;\pi\pi}^{IJ\,^*}\mathcal{M}_{a\pi;IJ}\,, \qquad  (E_{cm}>2m_\pi)\label{eq.imt}
\end{equation}
with 
\begin{equation}\label{eq.sigma}
\sigma_{\pi}(s)=\sqrt{1-\frac{4m_\pi^2}{s}}\,,
\end{equation}
and the BE distribution factor
\begin{equation}
n_B(E)= \frac{1}{e^{\frac{E}{T}}-1} \,. 
\end{equation}
In the zero-temperature limit, the BE distribution factor $n_B(E)$ tends to vanish and the thermal unitarity relation in Eq.~\eqref{eq.imt} recovers the standard one at zero temperature given in Ref.~\cite{DiLuzio:2022gsc}. A subtle difference comes up in the finite-temperature case that additional thermal Landau cuts generally appear~\cite{Weldon:1983jn,Ghosh:2010hap,GomezNicola:2023rqi}. E.g., in the $K\pi\to K\pi$ scattering process, the thermal Landau cuts generated in the $u$ channel show up in the physical energy region above the $K\pi$ threshold~\cite{GomezNicola:2023rqi}. In the current study of the $a\pi\to\pi\pi$ process, apart from the thermal unitarity cuts shown in Eq.~\eqref{eq.imt}, there are also thermal Landau cuts in the energy region $E_{cm}>2m_\pi$ contributed from both the $t$ and $u$ channels, which magnitudes turn out to be generally much smaller than the unitarity cuts.

\begin{figure}[t]
	\centering
	\includegraphics[angle=0.0,width=0.95\textwidth]{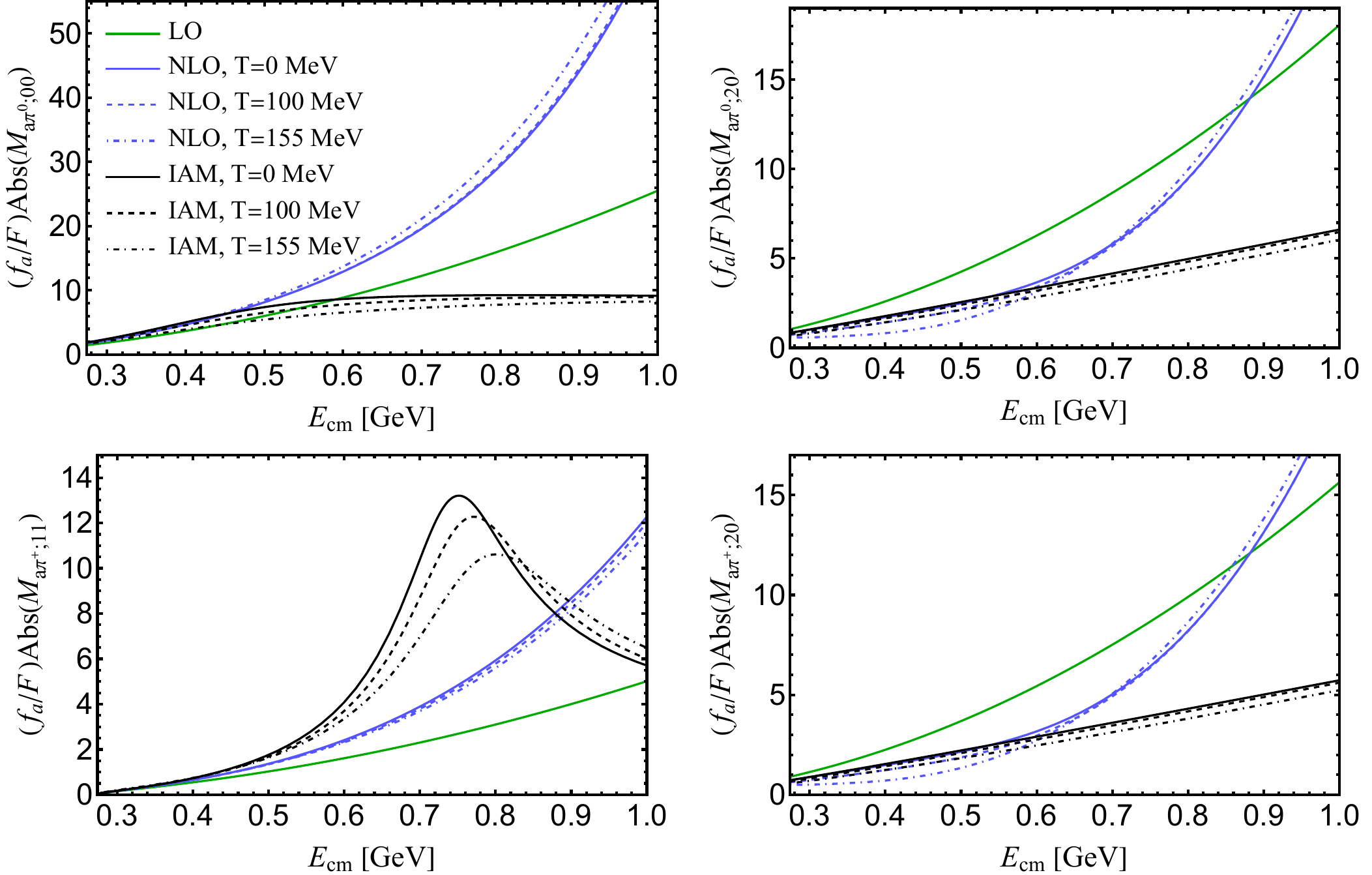}
	\caption{Magnitudes of the various $a\pi\to\pi\pi$ partial-wave amplitudes in the isospin bases. The curves labeled as NLO include both the contributions at $\mathcal{O}(p^2)$ and $\mathcal{O}(p^4)$. \label{fig.pwt}}
\end{figure}

\subsection{Surveys of the thermal corrections to the axion-pion scattering amplitudes and cross sections}

\begin{figure}[t]
	\centering
	\includegraphics[angle=0.0,width=1.0\textwidth]{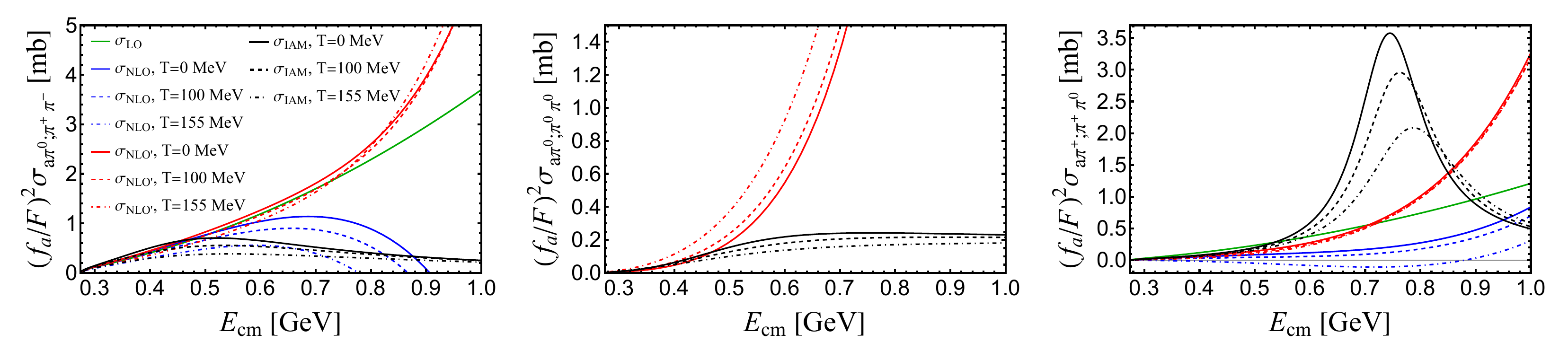}
	\caption{Cross sections in the charged bases. For the notaions of $\sigma_{\text{LO,NLO,NLO',IAM}}$, see the text for details. Notice that the LO contributions to the $a\pi^0\to \pi^0\pi^0$ amplitude and cross section vanish.  \label{fig.cscb}}
\end{figure}

In this part, we examine to what extent the finite-temperature corrections can affect the axion-pion scattering amplitudes and cross sections, comparing with the zero-temperature results. To make close comparison with the zero-temperature study of Ref.~\cite{DiLuzio:2022gsc}, we will take the same inputs for the $\mathcal{O}(p^4)$ LECs and other parameters as those used in the former reference, i.e., in perturbative calculation we take $\bar{l}_1= -0.36(59)$, $\bar{l}_2= 4.31(11)$, $\bar{l}_3= 3.53(26)$, $\bar{l}_4= 4.73(10)$, $l_7= 0.007(4)$, together with $m_u/m_d=0.50(2)$, $F_\pi=92.1(8)$~MeV and $m_\pi=137$~MeV; while for the IAM results we take the combinations $\bar{l}_1-\bar{l}_2=-5.95(2)$, $\bar{l}_1+\bar{l}_2=4.9(6)$, and others are same as the perturbative case.

In Fig.~\ref{fig.pwt}, we give the magnitudes of different PW amplitudes at $T=0$, 100 and 155~MeV. The first lesson we can learn from this figure is that the thermal corrections to the perturbative NLO amplitudes are insignificant up to $T=155$~MeV. The largest thermal effects appear in the IAM amplitudes for the $IJ=11$ channel, while the thermal corrections to the IAM amplitudes in other channels look small up to $T=155$~MeV. 

We then calculate the cross sections for the $a\pi\to\pi\pi$ in charged bases, both at zero and finite temperatures,
\begin{equation}\label{eq.cs}
\sigma(E_{cm})=\mathcal{N}\frac{1}{16(2\pi)^2E_{cm}^2}\frac{|\vec{p}_f|}{|\vec{p}_i|}\int\mathrm{d}\Omega\,
\left|\mathcal{M}\right|^2\,,
\end{equation}
where $|\vec{p}_i|$ and $|\vec{p}_f|$ are the magnitudes of three-momenta of the initial and final particles in the CM frame, and $\mathcal{N}=1/2$ for identical particles and otherwise $\mathcal{N}=1$. 
The comparisons are shown in Fig.~\ref{fig.cscb}. For the evaluation of the cross sections by using the perturbative amplitudes, one can expand the amplitude squared as 
\begin{equation}\label{eq.ampm2}
|\mathcal{M}|^2= |\mathcal{M}^{(2)}|^2+2\mathcal{M}^{(2)}{\rm Re}\big( \mathcal{M}^{(4)}\big)+|\mathcal{M}^{(4)}|^2 \,,
\end{equation}
where the property that the LO $\mathcal{M}^{(2)}$ is real has been used. 
In later discussions, we will denote $\sigma_{\rm LO}$ as the LO cross section when taking $|\mathcal{M}^{(2)}|^2$ in Eq.~\eqref{eq.ampm2}, and designate $\sigma_{\rm NLO}$ when taking the first two terms of Eq.~\eqref{eq.ampm2}, and $\sigma_{\rm  NLO'}$ when taking all the three terms of Eq.~\eqref{eq.ampm2} to evaluate the cross sections in Eq.~\eqref{eq.cs}. When using the unitarized IAM amplitude $\mathcal{M}^{\rm IAM}$ in Eq.~\eqref{eq.ampiam} to calculate the cross sections, the result is simply denoted as $\sigma_{\rm IAM}$. One can gain a rough estimate about how the next-to-next-to-leading order perturbative corrections may affect the cross sections by comparing $\sigma_{\rm NLO}$ with $\sigma_{\rm NLO'}$ at low energies.

The curves corresponding to the IAM amplitudes in Figs.~\ref{fig.pwt} and \ref{fig.cscb} are considered to be the preferred results in this work. By taking a closer look at different curves, one can gain a rough conclusion that the perturbative NLO amplitudes begin to be unreliable around $E\simeq0.5$~GeV, above which the next-to-next-to-leading order and IAM results begin to deviate noticeably.

\section{Brief discussions on the axion thermalization rate and its cosmology implication}\label{sec.cosmo}

We will closely follow Refs.~\cite{DiLuzio:2021vjd,DiLuzio:2022gsc} to proceed a brief discussion of the cosmological constraint on the axion parameters based on our updated calculation of the thermal $a\pi\to\pi\pi$ amplitudes. The axion decoupling temperature $T_D$ plays the key role, and according to the proposal in Ref.~\cite{Hannestad:2005df} it can be estimated as the temperature at which the axion interaction rate turns to be lower than the expansion rate of the universe. The explicit freeze out condition advocated in Ref.~\cite{Hannestad:2005df} to determine the axion decoupling temperature is 
\begin{equation}\label{eq.deftd}
\Gamma_a(T_D)=  H(T_D)\,,
\end{equation}
where $H(T)=T^2\sqrt{4\pi^3g_{*}(T)/45}/m_{\rm Pl}$ denotes the Hubble expansion parameter, with $g_{*}(T)$ the effective number of relativistic thermal degrees of freedom (d.o.f) and $m_{\rm Pl}$ the Plank mass.  
$\Gamma_a(T)$ in Eq.~\eqref{eq.deftd} stands for the averaged axion thermalization rate in the pion thermal bath, and it can be calculated via 
\begin{align}
\Gamma_a(T)=\frac{1}{n_a^{\text{eq}}}\int\mathrm{d}\widetilde{\Gamma}\,
\sum |\mathcal{M}_{a\pi;\pi\pi}|^2\,
n_B(E_1)n_B(E_2)[1+n_B(E_3)][1+n_B(E_4)]\,,\label{eq.gammat}
\end{align}
where the sum runs over all the possible $a\pi$ reaction channels, the axion number density in equilibrium is $n_a^{\text{eq}}=\zeta(3)\,T^3/\pi^2$ with $\zeta$ the Riemann Zeta function, and the phase space integral is 
\begin{equation}\label{eq.PSInt}
\int\mathrm{d}\widetilde{\Gamma}=
\int\left(\prod_{i=1}^4\frac{\mathrm{d}^3p_i}{(2\pi)^3}\frac{1}{2E_i}\right)
(2\pi)^4\delta^4(p_1+p_2-p_3-p_4)\,.
\end{equation} 
The recipe given in Ref~\cite{Green:2021hjh} will be used to calculate the above phase space integral. For the sake of completeness, we give some details about the evaluation of this integral in Appendix~\ref{appendix.D}. 
A more sophisticated way to solve the momentum dependent Boltzmann equations for the axion-pion scatterings, rather than taking the criteria in Eq.~\eqref{eq.deftd}, is proposed in Refs.~\cite{Notari:2022ffe,Bianchini:2023ubu}.   
However, the primary aim of this work is to examine to what extent the thermal corrections to the $a\pi\to\pi\pi$ scattering amplitudes can influence the determinations of the axion parameters, comparing with the situation by using the amplitudes at zero temperature, and therefore we will stick to the criteria in Eq.~\eqref{eq.deftd} to proceed.  

\begin{figure}[t]
	\centering
	\subfigure[\label{fig.hfunctions}]{
		\includegraphics[align=c,scale=0.28]{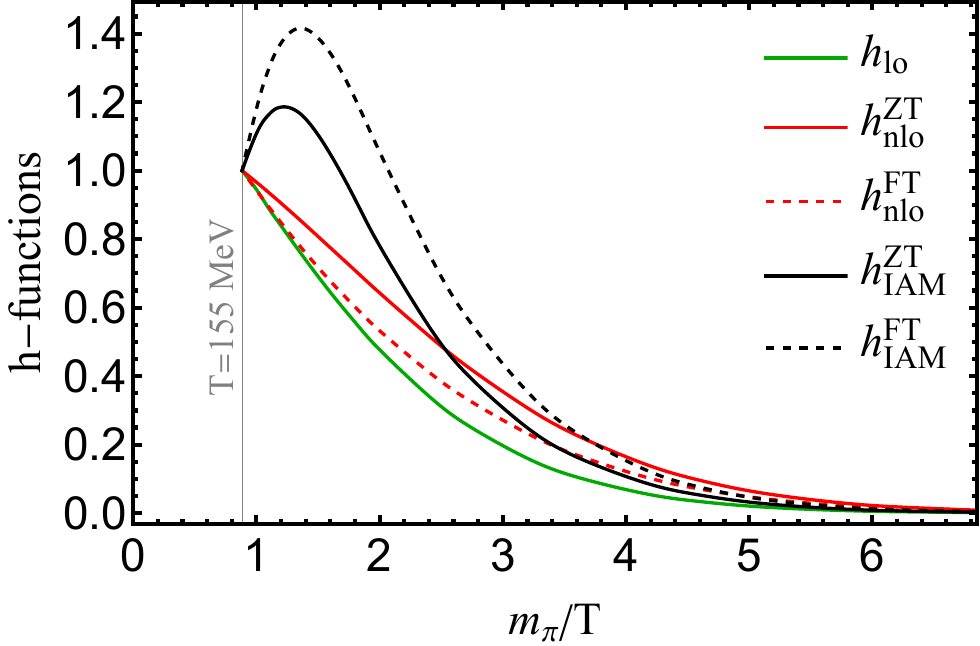}
	}
	\subfigure[\label{fig.RatePert}]{
		\includegraphics[align=c,scale=0.26]{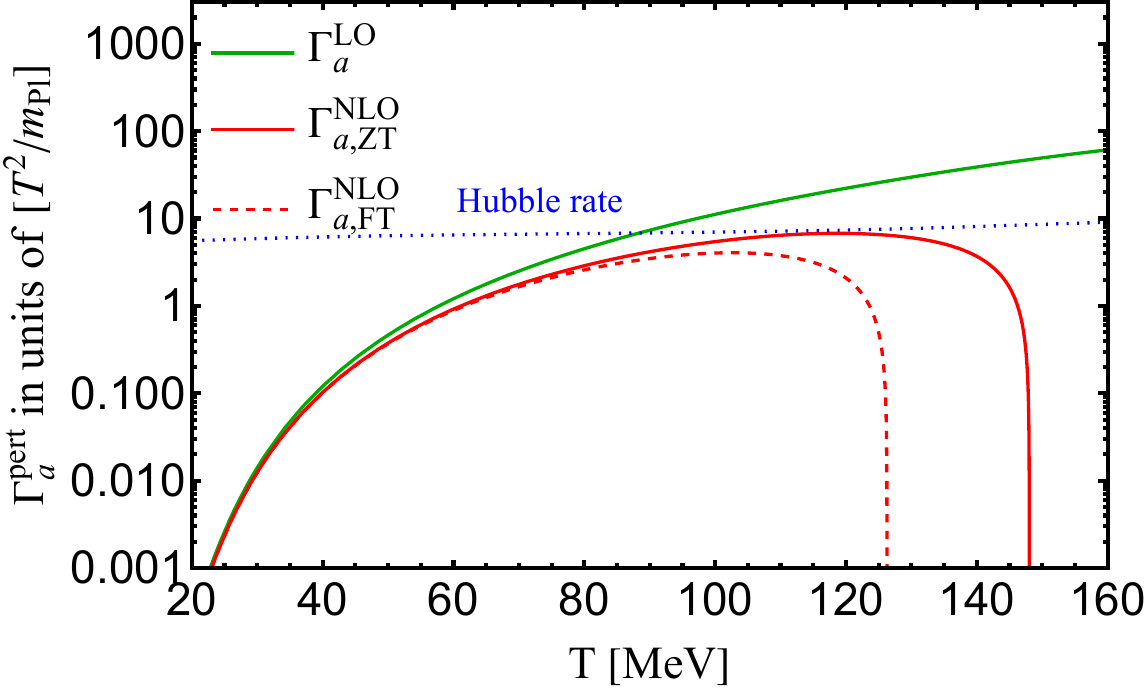}
	}
	\subfigure[\label{fig.rgamma}]{
		\includegraphics[align=c,scale=0.27]{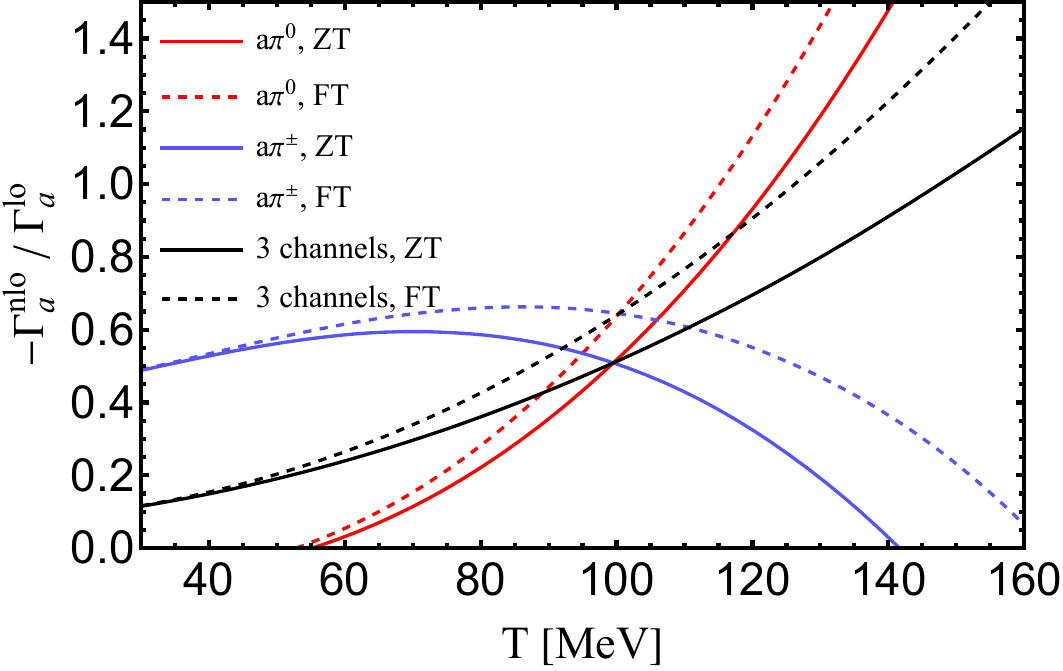}
	}
	\caption{  The symbols of ZT and FT indicate that the results are calculated by taking the zero-temperature and finite-temperature scattering amplitudes, respectively.
		(a) The curves for the $h_{\text{lo,\,nlo,\,IAM}}(m_\pi/T)$ functions are defined in Eqs.~\eqref{eq.hpert} and \eqref{eq.hIAM}.
		(b) The axion rates in this figure are calculated by taking the perturbative $a\pi\to\pi\pi$ scattering amplitudes upto NLO, with $f_a=1.5\times10^7$~GeV for illustration. $\Gamma_a^{\text{LO}}$ only includes the contribution from $h_{\text{lo}}$, and $\Gamma_a^{\text{NLO}}$ contains the contribution both from $h_{\text{lo}}$ and $h_{\text{nlo}}$. 
		(c) Ratio between the NLO parts and LO parts of the axion thermal averaged rates. $\Gamma_a^{\text{lo}}$ denotes the LO part of the perturbative rate and $\Gamma_a^{\text{nlo}}$ denotes the part of perturbative rate contributed only by $h_{\text{nlo}}$.
	}
\end{figure}

For the perturbative calculation up to NLO, we expand the squared amplitude as $\sum|\mathcal{M}|^2\simeq\sum|\mathcal{M}^{(2)}|^2+\sum 2\mathcal{M}^{(2)}\text{Re}(\mathcal{M}^{(4)})$, and then the perturbative axion rate will be casted into the following form~\cite{Hannestad:2005df}
\begin{equation}\label{eq.hpert}
\Gamma_a^{\text{pert}}(T)=
\left(\frac{\delta_I}{3f_aF}\right)^2T^5 \left[ \tilde{h}_{\rm lo} \, h_{\rm lo}\left(\frac{m_\pi}{T}\right) + \frac{T^2}{F^2} \tilde{h}_{\rm nlo}\, h_{\rm nlo}\left(\frac{m_\pi}{T} \right)\right]\,,
\end{equation}
where the dimensionless functions $h_{\text{lo}}$ and $h_{\text{nlo}}$ denote the contributions from $\sum|\mathcal{M}^{(2)}|^2$ and $\sum 2\text{Re}(\mathcal{M}^{(2)}\mathcal{M}^{(4)})$, respectively. The numeric factors $\tilde{h}_{\text{lo}}$ and $\tilde{h}_{\text{nlo}}$ are introduced so that $h_{\text{lo}}(m_\pi/T)$ and $h_{\text{nlo}}(m_\pi/T)$ are separately normalized at unity at the temperature $T_c$, which will be set as 155~MeV in order to make close comparisons with the results in Refs.~\cite{DiLuzio:2021vjd,DiLuzio:2022gsc}. The values of these factors are: $\tilde{h}_{\text{lo}}\simeq0.164$, $\tilde{h}_{\text{nlo}}\simeq-0.0550$ (when using zero-temperature amplitudes) and $-0.0755$ (when using thermal amplitudes).
When using IAM amplitudes as inputs, the rate is simply expressed as
\begin{equation}\label{eq.hIAM}
\Gamma_a^{\text{IAM}}(T)=\left(\frac{\delta_I}{3f_aF}\right)^2T^5\tilde{h}_{\text{IAM}}h_{\text{IAM}}\left(\frac{m_{\pi}}{T}\right)\,.
\end{equation}
For comparison, $h_{\text{IAM}}(m_{\pi}/T)$ is also normalized to unity at $T_c=155$~MeV by the numeric factor $\tilde{h}_{\text{IAM}}\simeq0.117$ (when using zero-temperature amplitudes) and $0.0817$ (when using thermal amplitudes). The results of $h_{\text{lo,nlo,IAM}}(m_{\pi}/T)$ are shown in Fig.~\ref{fig.hfunctions}.

For the axion thermalization rates estimated from the perturbative $\xpt$ amplitudes, the corresponding curves are given in Fig.~\ref{fig.RatePert}. After considering the NLO corrections of the amplitudes, the axion rate even drops to a negative value at high temperatures, due to the extreme breakdown of the chiral expansion in these regions. In Refs.~\cite{DiLuzio:2021vjd,DiLuzio:2022gsc}, it is proposed to inspect the valid region of $\xpt$ by using the ratios of the NLO and LO parts of the thermalization rates as a criteria. Our results for such ratios as a function of $T$ are given in Fig.~\ref{fig.rgamma}.
The breakdown temperature of $\xpt$ was extracted from the maximun of the ratio for the $a\pi^+$ channel~\cite{DiLuzio:2022gsc}. By approximating the $a\pi$ scattering amplitude from its zero-temperature expression, we confirm that the $\xpt$ breakdown temperature deduced from the $a\pi^+$ channel is around $70$~MeV, and at this point the ratio reaches the maximum value that is around $60\%$, see the blue solid line in Fig.~\ref{fig.rgamma}. For the result after taking the thermal correction to the $a\pi$ scattering amplitude, i.e. the blue dashed line, the maximum point of the ratio for $a\pi^+$ channel increases to around $85$~MeV, with the maximum ratio around $65\%$. The dashed lines in Fig.~\ref{fig.rgamma} are all above the solid lines, thus the convergence of the chiral expansion for axion thermalization rate gets worse after taking the thermal corrections into the $a\pi$ scattering amplitudes. Therefore after including the thermal corrections to the $a\pi\to\pi\pi$ amplitudes it is still unlikely to extract reliable bounds from the axion thermalization rates by using the perturbative $\xpt$ amplitudes, validating the previous conclusion in Refs.~\cite{DiLuzio:2021vjd,DiLuzio:2022gsc} that was obtained by taking the zero-temperature $a\pi\to\pi\pi$ amplitudes.

\begin{figure}[htb]
	\centering
	\subfigure[\label{fig.hubblegamma}]{
		\includegraphics[align=c,scale=0.38]{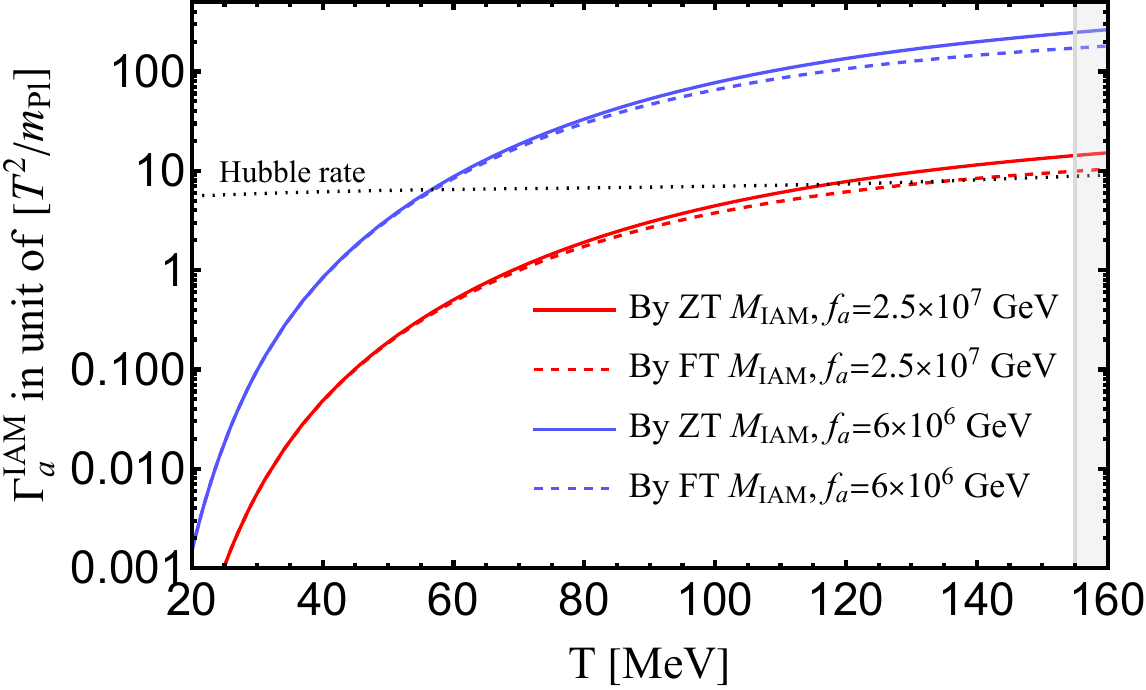}
	}
	\subfigure[\label{fig.TD-fa}]{
		\includegraphics[align=c,scale=0.36]{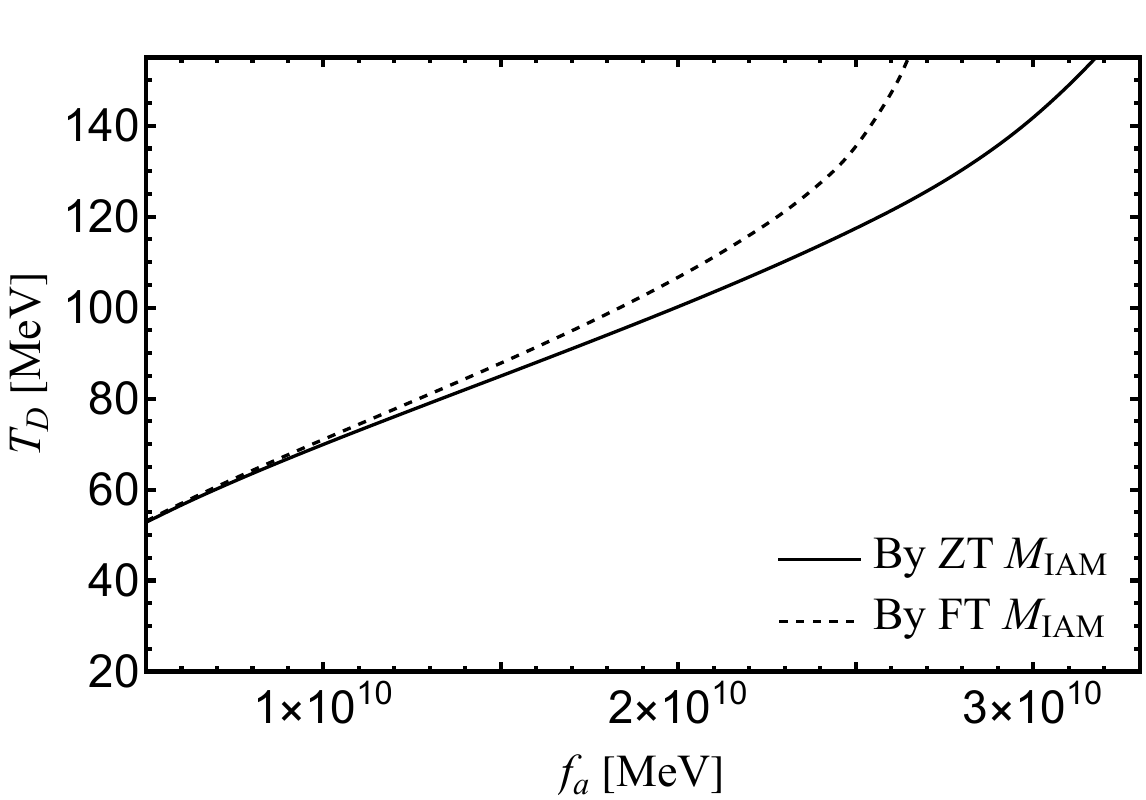}
	}
	\\
	\subfigure[\label{fig.DeltaNeff-fa}]{
		\includegraphics[align=c,scale=0.36]{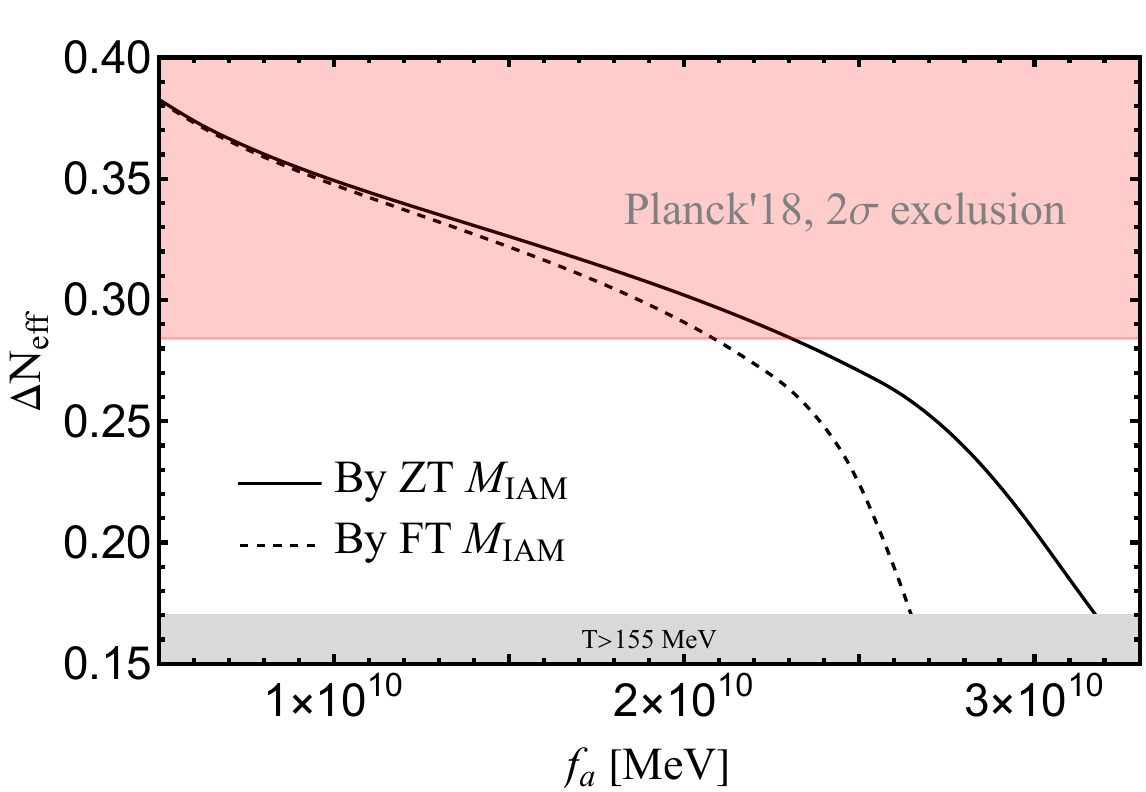}
	}
	\subfigure[\label{fig.maneff}]{
		\includegraphics[align=c,scale=0.36]{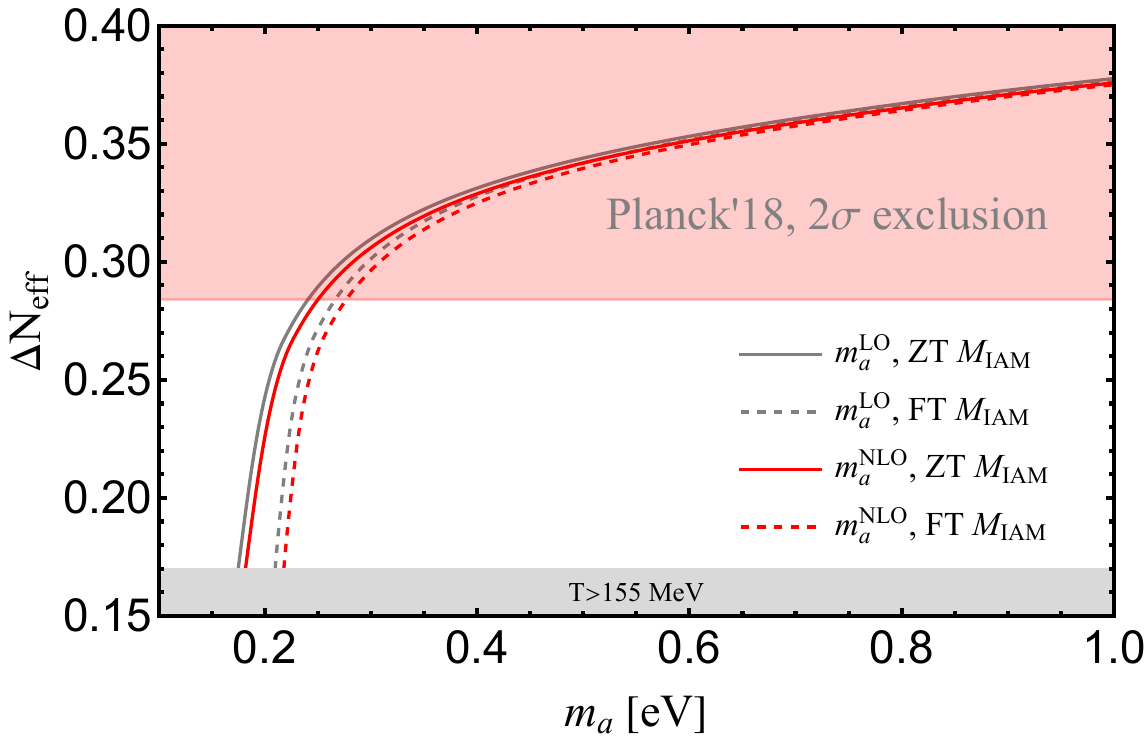}
	}
	\caption{ The symbols of ZT and FT indicate that the results are calculated by taking the zero-temperature and finite-temperature scattering amplitudes, respectively.
		(a) Axion thermalization rates calculated with the IAM amplitudes as a function of $T$. The solid lines denote the results by taking the amplitudes at zero temperature as an approximation and the dashed lines correspond to the updated results by including the thermal corrections to the $a\pi\to\pi\pi$ amplitudes.
		(b) Axion decoupling temperature as a function of the axion decay constant $f_a$. 
		(c) Extra effective thermal relativistic d.o.f. as a function of $f_a$.
		(d) The bounds on the axion masses $m_a$. The notations of $m_a^{\rm LO}$ and $m_a^{\rm NLO}$ refer to the situations by using the LO/NLO expressions for the axion masses.
	}
\end{figure}

The axion averaged thermalization rates as a function of $T$ from the IAM amplitudes are illustrated in Fig.~\ref{fig.hubblegamma}, where two groups of lines corresponding to two specific values of $f_a$ are given. 
We then compare the axion rates by scanning the values of $f_a$ with the Hubble parameters by taking the cosmological inputs from Ref.~\cite{Saikawa:2018rcs}. According to the criteria in Eq.~\eqref{eq.deftd}, the interception point of the two curves of $\Gamma_a(T)$ and $H(T)$ correspond to the axion decoupling temperature $T_D$, which varies with the value of $f_a$. In Fig.~\ref{fig.TD-fa}, we show the relations between $T_D$ and $f_a$ by taking the scattering amplitudes both at zero and finite temperatures to evaluate $\Gamma_a(T)$. The effects from the thermal corrections to the scattering amplitudes become noticeably important at higher values of $f_a$.

To take a comprehensive study of the cosmological analysis goes beyond the scope of this work, here we will concentrate on exploring the constraint of the extra effective number of relativistic d.o.f:   
\begin{eqnarray}
\Delta N_{\rm eff} \simeq \frac{4}{7} \left(\frac{43}{4g_{\star s}(T_D)}\right)^{\frac{4}{3}}\,,
\end{eqnarray}
on the axion parameters as done in Refs.~\cite{DiLuzio:2021vjd,DiLuzio:2022gsc}, where $g_{\star s}(T)$ corresponds to the effective number of entropy d.o.f at temperature $T$, and the cosmological determinations in Ref.~\cite{Saikawa:2018rcs} will be used in our analysis. The constraint of $\Delta N_{\rm eff}$ on the axion decay constant $f_a$ is given in  Fig.~\ref{fig.DeltaNeff-fa}, where the gray area in the bottom corresponds to the region for $T>155$~MeV and the IAM $\xpt$ is considered to be untrustworthy in this region. Our study indicates that the inclusion of the thermal corrections to the scattering amplitudes can slightly lower the bound  $f_a\gtrsim 2.1\times 10^7$~GeV, comparing with the constraint $f_a\gtrsim 2.3\times 10^7$~GeV obtained by taking the $a\pi\to\pi\pi$ scatering amplitudes at zero temperature. That is to say the thermal corrections to the scattering amplitudes can cause around a 10\% shift to the constraint of the axion decay constant $f_a$.  

In turn one could translate the bounds of $f_a$ into the constraints of the axion mass $m_a$, relying on the LO/NLO $\xpt$ predictions in Appendix~\ref{appendix.B}.  The constraints on the QCD axion mass $m_a$ are shown in Fig.~\ref{fig.maneff}, where we have distinguished the constraints by including/excluding the thermal corrections to the $a\pi$ scattering amplitudes and also using the LO/NLO expressions for the axion masses.
When excluding the thermal corrections of $a\pi$ scattering amplitudes, the bounds for the axion masses are $m_a\lesssim 0.24$~eV by $m_a^{\text{LO}}$ and $m_a\lesssim 0.25$~eV by $m_a^{\text{NLO}}$\,.
When including the thermal corrections of $a\pi$ scattering amplitudes, the bounds for the axion mass are $m_a\lesssim 0.27$~eV by $m_a^{\text{LO}}$ and $m_a\lesssim 0.28$~eV by $m_a^{\text{NLO}}$\,.
Therefore the thermal corrections of the amplitudes introduce a $10\%$ variation to the determination of the upper limits of $m_a$.

\section{Summary and conclusions}\label{sec.concl}

In this work we have performed the complete calculation of the $a\pi\to\pi\pi$ scattering amplitudes up to the one-loop level at finite temperatures within the $SU(2)$ QCD axion chiral perturbation theory. The inverse amplitude method is further used to unitarize the partial-wave $a\pi\to\pi\pi$ scattering amplitudes, which extends the applicable energy and temperature regions of the amplitudes in chiral perturbation theory. The largest effect from the thermal correction to the $a\pi\to\pi\pi$ scattering amplitudes appears in the $a\pi^{\pm}\to \pi^{\pm}\pi^0$ channels with $IJ=11$ for the $\pi\pi$ system, while the thermal effects in other scattering amplitudes are small. 

The axion averaged thermalization rates are then calculated by using the updated $a\pi\to\pi\pi$ scattering amplitudes at finite temperatures. The axion decoupling temperatures are obtained with the averaged axion thermalization rates calculated from the thermal $a\pi\to\pi\pi$ scattering amplitudes within the inverse amplitude method. The extra effective number of relativistic degrees of freedom $\Delta N_{\rm eff}$ is then used to constrain the axion decay constant $f_a$, which gives $f_a>2.1 \times 10^7$~GeV, comparing with the bound of $f_a>2.3 \times 10^7$~GeV when neglecting the thermal corrections in the $a\pi$ scattering amplitudes. This 10\% correction effect for the lower limit of $f_a$ also delivers a similar magnitude of shift to the upper limit of the axion mass. 
Therefore our study shows that the inclusion of the thermal effects in the $a\pi\to\pi\pi$ scattering amplitudes can cause a mild shift (around 10\%) of the axion parameters when confronting with the cosmological constraint of $\Delta N_{\rm eff}$. It is necessary to include the thermal corrections in the $a\pi$ scattering amplitudes in the future attempts that aim to improve the cosmological determination of the axion parameters at this level.

\section*{Acknowledgements}

We would like to thank Yue Zhang for reading the manuscript and suggestions. We also thank Luca Di Luzio, Jos\'e Antonio Oller and Gioacchino Piazza for clarifying the axion-pion amplitudes at zero temperature.  
This work is funded in part by the National Natural Science Foundation of China (NSFC) under Grants Nos.~12150013, 11975090, and the Science Foundation of Hebei Normal University with contract No.~L2023B09.  

\appendix

\section{Relevant one-loop integrals at finite temperatures}\label{appendix.A}
\setcounter{equation}{0}
\def\theequation{A.\arabic{equation}}

In the imaginary time formalism, the finite-temperature effects enter via the chiral loops by replacing the temporal integrals with the Matsubara sums, i.e.,
\begin{equation}
-i\int\frac{\mathrm{d}^dq}{(2\pi)^d}\,I(q^0,\,\vec{q}\,)\rightarrow
-i\int_{\beta}\frac{\mathrm{d}^dq}{(2\pi)^d}\,I(q^0,\,\vec{q}\,)= 
T\sum_{n}\int\frac{\mathrm{d}^{d-1}q}{(2\pi)^{d-1}}\,I(i\omega_n,\,\vec{q}\,)\,,
\end{equation}
where the replacement of $q^0\rightarrow{i\omega_n}=i2\pi{nT}$ with $n\in{\mathbb{Z}}$ is taken and the symbol $\beta$ is introduced to highlight the integrals with finite-temperature corrections. Meanwhile, the zeroth components of all external momenta also need to be replaced by the Matsubara frequencies, which can be extrapolated to real energies after completing the Matsubara sums. When performing such sums, one can always separate the results into the zero-temperature part and the finite-temperature correction part: 
\begin{align}
&G(T)= T\sum_{n}\int\frac{\mathrm{d}^{d-1}q}{(2\pi)^{d-1}}\,I(i\omega_n,\,\vec{q}\,)\notag 
\\
=&
\int\frac{\mathrm{d}^dq_E}{(2\pi)^d}\,I(iq_E^0,\,\vec{q}\,)
+\int_{-\infty-i0^+}^{+\infty-i0^+}\frac{\mathrm{d}z_E}{2\pi}\int\frac{\mathrm{d}^3q}{(2\pi)^3}
\left[I(iz_E,\,\vec{q}\,)+I(-iz_E,\,\vec{q}\,)\right]n_B(iz_E) 
\notag \\
\equiv& G(T=0) + \Delta G (T)\,,
\label{EQ-A2}
\end{align}
where the first term corresponds to the zero-temperature integral and the second one stands for the finite-temperature correction. One can use the standard dimensional regularization to calculate the zero-temperature integral, in which the UV divergence will appear. We use the conventional $\overline{MS}-1$ renormalization scheme that is widely employed in $\xpt$~\cite{Gasser:1983yg} for the zero-temperature integrals $G(T=0)$.

Next we give relevant formulas for the one-loop integrals at finite temperatures, i.e. the $\Delta G(T)$ part. The finite-temperature integrals in this work are the same as those in thermal $\pi\pi$ scattering~\cite{GomezNicola:2002tn}, which include the tadpole integrals, like the diagram~\ref{fig.1c}, and the one-loop two-point integrals for diagram~\ref{fig.1d}-\ref{fig.1f}.

\subsection{Tadpole loop integral}

According to Eq.~\eqref{EQ-A2}, the basic tadpole integral reads  
\begin{equation}\label{eq.defintf}
F_{\beta}(T)\equiv-i\int_{\beta}\frac{\mathrm{d}^dq}{(2\pi)^d}\frac{1}{q^2-m^2}=F_\beta(0)+\Delta F_\beta(T)\,,
\end{equation}
where the zero-temperature integral in the $\overline{MS}-1$ scheme is 
\begin{equation}
F_{\beta}(0)=-\frac{m^2}{16\pi^2}\log{\frac{m^2}{\mu^2}}\,,
\end{equation}
and the finite-temperature correction is 
\begin{equation}
{\Delta}F_{\beta}(T)=-\frac{1}{2\pi^2}\int_m^{+\infty}\mathrm{d}E_q\sqrt{E_q^2-m^2}\, n_B(E_q)\,,
\end{equation}
with $E_q=\sqrt{|\vec{q}|^2-m^2}$. 

For the integrals with $p \cdot q$ or $q^2$ in the numerators, the results are 
\begin{align}
&-i\int_{\beta}\frac{\mathrm{d}^dq}{(2\pi)^d}\frac{p \cdot q}{q^2-m^2}=0\,,
\\
&-i\int_{\beta}\frac{\mathrm{d}^dq}{(2\pi)^d}\frac{q^2}{q^2-m^2}=m^2F_{\beta}(T)\,.
\end{align}

\subsection{Basic two-point one-loop integrals at finite temperatures}

Four types of the two-point one-loop integrals will appear in the calculation of $\xpt$ 
\begin{eqnarray}\label{eq.def4int}
-i\int_{\beta}\frac{\mathrm{d}^dq}{(2\pi)^d}
\frac{(1, \,\, p_1 \cdot q, \,\, q^2,\,\, p_1\cdot q p_2\cdot q)}{(q^2-m^2)[(k-q)^2-m^2]}\,,
\end{eqnarray}
whose zero-temperature parts can be given in terms of the $B_0(k^2,m^2)$ function~\cite{Gasser:1983yg} 
\begin{eqnarray}
16\pi^2 B_0(k^2,m^2)= -\sigma_m\log\frac{\sigma_m-1}{\sigma_m+1}-1+\log{\frac{m^2}{\mu^2}}\,,\quad
\sigma_m(k^2)=\sqrt{1-\frac{4m^2}{k^2}}\,,
\end{eqnarray}
due to the Lorentz invariance. In the following, we refrain from discussing the zero-temperature parts of the integrals in Eq.~\eqref{eq.def4int} and focus on the finite-temperature correction ($\mathcal{FTC}$). While for the latter part, the Lorentz invariance is lost and one needs to separately treat the temporal and spatial components of the integration variable $q^\mu=(q^0,q^i)$ in the evaluation.  It turns out that the $\mathcal{FTC}$ parts of the four types of integrals in Eq.~\eqref{eq.def4int} can be expressed in terms of three independent thermal loop functions~\cite{GomezNicola:2002tn}
\begin{equation}\label{eq.defintjl}
\Delta J_l(T,k^0,|\vec{k}|)\equiv-i\int_{\beta}\frac{\mathrm{d}^dq}{(2\pi)^d}
\frac{{(q^0)}^l}{(q^2-m^2)[(k-q)^2-m^2]}\bigg|_{\mathcal{FTC}}\,,  \qquad  (l=0,1,2)\,,
\end{equation}
where the subscript $\mathcal{FTC}$ is introduced to emphasize that $\Delta J_l(T,k^0,|\vec{k}|)$ only include the finite-temperature parts of the integrals.   
After completing the Matsubara sums, the integrals in Eq.~\eqref{eq.defintjl} can be written as
\begin{equation}
{\Delta}J_l(T,k^0,|\vec{k}|)=\sum_{\lambda_1,\lambda_2=\pm 1}f_{\lambda_1\lambda_2}^{(l)}(T,k^0,|\vec{k}|)\,,
\end{equation}
with 
\begin{equation}\label{eq.f-function}
\begin{aligned}
f_{\lambda_1\lambda_2}^{(l)}(T,k^0,|\vec{k}|)=&\int\frac{\mathrm{d}^3q}{(2\pi)^3}\frac{1}{4E_qE_{k-q}}
\frac{\lambda_1\lambda_2}{k^0-\lambda_1E_q-\lambda_2E_{k-q}+i0^+}
\\
&\times\left[-\lambda_1(\lambda_1E_q)^ln_B(E_q)-\lambda_2(k^0-\lambda_2E_{k-q})^ln_B(E_{k-q})\right]\,,
\end{aligned}
\end{equation}
and $E_{k-q}=\sqrt{(\vec{k}-\vec{q}\,)^2+m^2}$\,. The functions $f_{\lambda_1\lambda_2}^{(l)}(T,k^0,|\vec{k}|)$ depend on $k^0$ and $|\vec{k}|$ in a separate manner. Before giving the technical details about the calculation of the thermal loop functions in Eq.~\eqref{eq.f-function}, we first elaborate the reduction formulas for the temperature correction parts of the two-point one-loop integrals in Eq.~\eqref{eq.def4int}.

\subsection{Reduction of the two-point one-loop integrals at finite temperatures}

The numerators with spatial components $q^i$ in the integrals of Eq.~\eqref{eq.def4int} can be written in terms of the basic ones in Eqs.~\eqref{eq.defintf} and \eqref{eq.defintjl}. For the sake of completeness, we give a brief discussion about the finite-temperature parts of the various integrals in Eq.~\eqref{eq.def4int}. 

For the three types of numerators with $(p_1 \cdot q),\,q^2,\,(p_1 \cdot q)(p_2 \cdot q)$ in the integrals of Eq.~\eqref{eq.def4int}, their $\mathcal{FTC}$ parts can be reduced to ${\Delta}F_{\beta}(T)$~\eqref{eq.defintf} and ${\Delta}J_{l}(T,k^0,|\vec{k}|)$~\eqref{eq.defintjl}. The $\mathcal{FTC}$ reduction formulas take the form

\vspace{2mm}
\noindent$\bullet~(p_1 \cdot q)$
\begin{equation}\label{eq.intfti}
\left.-i\int_{\beta}\frac{\mathrm{d}^dq}{(2\pi)^d}
\frac{p_1 \cdot q}{(q^2-m^2)[(k-q)^2-m^2]}\right|_{\mathcal{FTC}}
=p_1^0\Delta J_1(T,k^0,|\vec{k}|)-(\vec{p}_1\cdot\vec{k})\Delta b_1(T,k^0,|\vec{k}|)\,,
\end{equation}
with
\begin{equation}
\Delta b_1(T,k^0,|\vec{k}|)=\frac{1}{2|\vec{k}|^2}\left[2k^0\Delta J_1(T,k^0,|\vec{k}|)-k^2\Delta J_0(T,k^0,|\vec{k}|)\right]\,;
\end{equation}

\vspace{2mm}
\noindent$\bullet~q^2$
\begin{equation}
\left.-i\int_{\beta}\frac{\mathrm{d}^dq}{(2\pi)^d}
\frac{q^2}{(q^2-m^2)[(k-q)^2-m^2]}\right|_{\mathcal{FTC}}
={\Delta}F_{\beta}(T)+m^2{\Delta}J_0(T,k^0,|\vec{k}|)\,;
\end{equation}

\vspace{2mm}
\noindent$\bullet~(p_1 \cdot q)(p_2 \cdot q)$
\begin{equation}
\begin{aligned}
&\left.-i\int_{\beta}\frac{\mathrm{d}^dq}{(2\pi)^d}
\frac{(p_1 \cdot q)(p_2 \cdot q)}{(q^2-m^2)[(k-q)^2-m^2]}\right|_{\mathcal{FTC}}
\\
=&p_1^0p_2^0\Delta J_2(T,k^0,|\vec{k}|)
-[p_1^0(\vec{p}_2\cdot\vec{k})+p_2^0(\vec{p}_1\cdot\vec{k})]\Delta b_{J1}(T,k^0,|\vec{k}|)
\\
&+(\vec{p}_1\cdot\vec{p}_2)\Delta b_{21}(T,k^0,|\vec{k}|)
+(\vec{p}_1\cdot\vec{k})(\vec{p}_2\cdot\vec{k})\Delta b_{20}(T,k^0,|\vec{k}|)\,,
\end{aligned}
\end{equation}
with
\begin{align}
&\Delta b_{J1}(T,k^0,|\vec{k}|)=\frac{1}{2|\vec{k}|^2}
\left[-k^0\Delta F_{\beta}(T)+2k^0\Delta J_2(T,k^0,|\vec{k}|)-k^2\Delta J_1(T,k^0,|\vec{k}|)\right]\,,
\\
&
\begin{aligned}
\Delta b_{21}(T,k^0,|\vec{k}|)=&\frac{1}{2|\vec{k}|^2}
\bigg[
\frac{k^2}{2}\Delta F_{\beta}(T)
-\frac{k^4+4m^2|\vec{k}|^2}{4}\Delta J_0(T,k^0,|\vec{k}|)
\\
&+k^0k^2\Delta J_1(T,k^0,|\vec{k}|)-k^2\Delta J_2(T,k^0,|\vec{k}|)
\bigg]\,,
\end{aligned}
\\
&
\begin{aligned}\label{eq.intftf}
\Delta b_{20}(T,k^0,|\vec{k}|)=&\frac{1}{2|\vec{k}|^4}
\bigg[
-\frac{3{k^0}^2+|\vec{k}|^2}{2}\Delta F_{\beta}(T)
+\frac{3k^4+4m^2|\vec{k}|^2}{4}\Delta J_0(T,k^0,|\vec{k}|)
\\
&-3k^0k^2\Delta J_1(T,k^0,|\vec{k}|)
+\left(3{k^0}^2-|\vec{k}|^2\right)\Delta J_2(T,k^0,|\vec{k}|)\bigg]\,.
\end{aligned}
\end{align}
It is reiterated that the above reduction formulas with the subscript of $\mathcal{FTC}$ include only the finite-temperature correction parts.

\subsection{Technical details about the evaluation of basic thermal loop functions}

We give some technical details below about how to evaluate the integrals in Eq.~\eqref{eq.f-function}. 
It is noted that $f_{\lambda_1\lambda_2}^{(l)}(T,k^0,|\vec{k}|)$ satisfy the following reflection relations with respect to $k^0$~\cite{Ghosh:2010hap}
\begin{equation}
\begin{aligned}
&f_{++}^{(0,2)}(T,-k^0,|\vec{k}|)=\left[f_{--}^{(0,2)}(T,k^0,|\vec{k}|)\right]^{*}\,,
\quad
f_{+-}^{(0,2)}(T,-k^0,|\vec{k}|)=\left[f_{-+}^{(0,2)}(T,k^0,|\vec{k}|)\right]^{*}\,,
\\
&f_{++}^{(1)}(T,-k^0,|\vec{k}|)=-\left[f_{--}^{(1)}(T,k^0,|\vec{k}|)\right]^{*}\,,
\quad
f_{+-}^{(1)}(T,-k^0,|\vec{k}|)=-\left[f_{-+}^{(1)}(T,k^0,|\vec{k}|)\right]^{*}\,,
\end{aligned}
\end{equation}
which lead to the following relations of the various loop functions introduced previously
\begin{align}
&{\Delta}J_{0,2}(T,-k^0,|\vec{k}|)=\left[{\Delta}J_{0,2}(T,k^0,|\vec{k}|)\right]^*\,,
\\
&{\Delta}J_1(T,-k^0,|\vec{k}|)=-\left[{\Delta}J_1(T,k^0,|\vec{k}|)\right]^*\,,
\\
&{\Delta}b_{(1,20,21)}(T,-k^0,|\vec{k}|)=\left[{\Delta}b_{(1,20,21)}(T,k^0,|\vec{k}|)\right]^*\,,
\\
&{\Delta}b_{J1}(T,-k^0,|\vec{k}|)=-\left[{\Delta}b_{J1}(T,k^0,|\vec{k}|)\right]^*\,.
\end{align}

Next we use a similar procedure in Ref.~\cite{Ghosh:2010hap} to simplify the expressions of $f_{\lambda_1\lambda_2}^{(l)}(T,k^0,|\vec{k}|)$ given in Eq.~(\ref{eq.f-function}). 
To proceed the evaluation of the integrals in Eq.~\eqref{eq.f-function}, we take $E_q=\sqrt{\vec{q}^2+m^2}$ and $E_{k-q}=\sqrt{\vec{k}^2+\vec{q}\,^2-2|\vec{k}||\vec{q}\,|\cos\theta+m^2}$ as the two integral variables, i.e.,
\begin{equation}
\int\frac{\mathrm{d}^3q}{(2\pi)^3}\frac{1}{4E_qE_{k-q}}
=\frac{1}{4(2\pi)^2|\vec{k}|}\int_m^{+\infty}\mathrm{d}E_q\int_{E_{k-q}^{\text{min}}}^{E_{k-q}^{\text{max}}}\mathrm{d}E_{k-q}\,,
\end{equation}
with
\begin{equation}
E_{k-q}^{\text{min/max}}=\sqrt{|\vec{k}|^2+E_q^2\mp2|\vec{k}|\sqrt{E_q^2-m^2}}\,.
\end{equation}
The real and imaginary parts of the $f_{\lambda_1\lambda_2}^{(l)}(T,k^0,|\vec{k}|)$ functions are given by 
\begin{align}
&\begin{aligned}
\text{Re}f_{\lambda_1\lambda_2}^{(l)}(T,k^0,|\vec{k}|)=&
\int_m^{+\infty}\mathrm{d}E_q\,\mathbb{P}\int_{E_{k-q}^{\text{min}}}^{E_{k-q}^{\text{max}}}\mathrm{d}E_{k-q}\frac{1}{4(2\pi)^2|\vec{k}|}
\frac{\lambda_1\lambda_2}{k^0-\lambda_1E_q-\lambda_2E_{k-q}}
\\
&\times\left[-\lambda_1(\lambda_1E_q)^ln_B(E_q)-\lambda_2(k^0-\lambda_2E_{k-q})^ln_B(E_{k-q})\right]\,,
\end{aligned}
\\
&\begin{aligned}
\text{Im}f_{\lambda_1\lambda_2}^{(l)}(T,k^0,|\vec{k}|)=&
-\frac{\pi}{4(2\pi)^2|\vec{k}|}\int_m^{+\infty}\mathrm{d}E_q\int_{E_{k-q}^{\text{min}}}^{E_{k-q}^{\text{max}}}\mathrm{d}E_{k-q}\delta(k^0-\lambda_1E_q-\lambda_2E_{k-q})
\\
&\times\lambda_1\lambda_2\left[-\lambda_1(\lambda_1E_q)^ln_B(E_q)-\lambda_2(k^0-\lambda_2E_{k-q})^ln_B(E_{k-q})\right]\,,
\end{aligned}
\end{align}
where $\mathbb{P}$ denotes taking principle value of the integrals. 
The $\delta$ functions in $\mathrm{Im}f_{++}^{(l)}(T,k^0,|\vec{k}|)$ and $\mathrm{Im}f_{--}^{(l)}(T,k^0,|\vec{k}|)$ are non-vanishing for $k^2>4m^2$\,, which give the unitary cuts. While the $\delta$ functions in $\mathrm{Im}f_{+-}^{(l)}(T,k^0,|\vec{k}|)$ and $\mathrm{Im}f_{-+}^{(l)}(T,k^0,|\vec{k}|)$ are non-vanishing for $k^2<0$\,, leading to the so-called Landau cuts.

Next we can integrate out $E_{k-q}$ in $\mathrm{Im}f_{\lambda_1\lambda_2}^{(l)}(T,k^0,|\vec{k}|)$ by the $\delta$ function. We take $\mathrm{Im}f_{++}^{(l)}(T,k^0,|\vec{k}|)$ for illustration. The integral for $\mathrm{Im}f_{++}^{(l)}(T,k^0,|\vec{k}|)$ is non-vanishing only if $E_{k-q}^{\text{min}}\leq k^0-E_q\leq E_{k-q}^{\text{max}}$, and to combine with $k^2>4m^2$ one can get
\begin{equation}
\frac{k^0}{2}-\frac{|\vec{k}|}{2}\sigma_m(k^2)\leq E_q\leq \frac{k^0}{2}+\frac{|\vec{k}|}{2}\sigma_m(k^2)\,. 
\end{equation}
Therefore, we have
\begin{align}
&\begin{aligned}
&\text{Im}f_{++}^{(l)}(T,k^0,|\vec{k}|)=
\frac{-\theta(k^0-E_k^{th})}{16\pi|\vec{k}|}
\int_{E_{q,++}^{\text{min}}}^{E_{q,++}^{\text{max}}}
\mathrm{d}E_q\,(E_q)^l\left[-n_B(E_q)-n_B(k^0-E_q)\right]\,,
\\
&E_{q,++}^{\text{min}}=\frac{k^0}{2}-\frac{|\vec{k}|}{2}\sigma_m(k^2)\,,
\quad
E_{q,++}^{\text{max}}=\frac{k^0}{2}+\frac{|\vec{k}|}{2}\sigma_m(k^2)\,,
\end{aligned}
\end{align}
with $E_k^{th}=\sqrt{|\vec{k}|^2+4m^2}$\,. Similarly, one can obtain
\begin{align}
&
\begin{aligned}
&\text{Im}f_{+-}^{(l)}(T,k^0,|\vec{k}|)=
\frac{-\theta(-k^2)}{16\pi|\vec{k}|}
\int^{+\infty}_{E_{q,+-}^{\text{min}}}
\mathrm{d}E_q\,(E_q)^l\left[n_B(E_q)-n_B(E_q-k^0)\right]\,,
\\
&E_{q,+-}^{\text{min}}=\frac{k^0}{2}+\frac{|\vec{k}|}{2}\sigma_m(k^2)\,;
\end{aligned}
\\
&
\begin{aligned}
&\text{Im}f_{-+}^{(l)}(T,k^0,|\vec{k}|)=
\frac{-\theta(-k^2)}{16\pi|\vec{k}|}
\int^{+\infty}_{E_{q,-+}^{\text{min}}}
\mathrm{d}E_q\,(-E_q)^l\left[-n_B(E_q)+n_B(k^0+E_q)\right]\,,
\\
&E_{q,-+}^{\text{min}}=-\frac{k^0}{2}+\frac{|\vec{k}|}{2}\sigma_m(k^2)\,;
\end{aligned}
\\
&
\begin{aligned}
&\text{Im}f_{--}^{(l)}(T,k^0,|\vec{k}|)=
\frac{-\theta(-k^0-E_k^{th})}{16\pi|\vec{k}|}
\int_{E_{q,--}^{\text{min}}}^{E_{q,--}^{\text{max}}}
\mathrm{d}E_q\,(-E_q)^l\left[n_B(E_q)+n_B(-k^0-E_q)\right]\,,
\\
&E_{q,--}^{\text{min}}=-\frac{k^0}{2}-\frac{|\vec{k}|}{2}\sigma_m(k^2)\,,
\quad
E_{q,--}^{\text{max}}=-\frac{k^0}{2}+\frac{|\vec{k}|}{2}\sigma_m(k^2)\,.
\end{aligned}
\end{align}
The remaining integration of $E_q$ in $\mathrm{Im}f_{\lambda_1\lambda_2}^{(l)}(T,k^0,|\vec{k}|)$ is straightforward and can be even analytically evaluated , nevertheless we do not explicitly show the results here, due to the rather lengthy expressions.

In the special case of $|\vec{k}|=0$\,, ${\Delta}J_1(T,k^0,0)$ and ${\Delta}J_2(T,k^0,0)$ can be reduced to ${\Delta}F_{\beta}(T)$ and ${\Delta}J_0(T,k^0,0)$: 
\begin{align}
&{\Delta}J_1(T,k^0,0)=\frac{k^0}{2}{\Delta}J_0(T,k^0,0)\,,
\\
&{\Delta}J_2(T,k^0,0)=\frac{1}{2}{\Delta}F_{\beta}(T)+\left(\frac{k^0}{2}\right)^2 \Delta J_0(T,k^0,0)\,,
\end{align}
with 
\begin{equation}
\begin{aligned}
{\Delta}J_0(T,k^0,0)= 
&\,\theta(k^0-2m)\frac{i\sigma_m({k^0}^2)}{8\pi}n_B\left(\frac{k^0}{2}\right)
-\theta(-k^0-2m)\frac{i\sigma_m({k^0}^2)}{8\pi}n_B\left(-\frac{k^0}{2}\right)
\\ 
&+\frac{1}{(2\pi)^2}\mathbb{P}\int_m^{+\infty}\mathrm{d}E_q\sqrt{1-\frac{m^2}{E_q^2}}
\left(\frac{1}{k^0+2E_q}-\frac{1}{k^0-2E_q}\right)n_B(E_q)\,. 
\end{aligned}
\end{equation}

\section{Two-point functions and masses of axion and pion at finite temperatures}\label{appendix.B}
\setcounter{equation}{0}
\def\theequation{B.\arabic{equation}}

The bilinear terms of $a$ and $\pi^0$ from the LO Lagrangian~\eqref{eq.L2} read 
\begin{equation}\label{EQ-B1}
\mathcal{L}_{2,bilinear}=
\frac{1}{2}\partial_{\mu}a\partial^{\mu}a+\frac{1}{2}\partial_{\mu}\pi^0\partial^{\mu}\pi^0
+\delta_{a\pi}\partial_{\mu}a\partial^{\mu}\pi^0
-\frac{1}{2}\bar{m}_a^2a^2-\frac{1}{2}\bar{m}_{\pi}^2\pi^0\pi^0\,,
\end{equation}
where $\delta_{a\pi}$ is given in Eq.~\eqref{eq.11} and the QCD axion mass at LO is 
\begin{equation}
\bar{m}_a^2=\gamma_{ud}\,\bar{m}_{\pi}^2\frac{F^2}{f_a^2}\,,
\qquad
\gamma_{ud}=\frac{m_um_d}{(m_u+m_d)^2}\,. 
\end{equation}
The barred quantities correspond to their LO expressions.  

Up to $\mathcal{O}(1/f_a^2)$\,, the $a$-$\pi^0$ mixing in Eq.~(\ref{EQ-B1}) can be eliminated by the field redefinition
\begin{equation}
a \to a-\delta_{a\pi}\pi^0+\mathcal{O}(1/f_a^3)\,,
\quad
\pi^0 \to \left(1+\frac{1}{2}\delta_{a\pi}^2\right)\pi^0+\mathcal{O}(1/f_a^3)\,,
\end{equation}
and the Lagrangian in Eq.~\eqref{EQ-B1} becomes
\begin{equation}
\mathcal{L}_{2,bilinear}=
\frac{1}{2}\partial_{\mu}a\partial^{\mu}a+\frac{1}{2}\partial_{\mu}\pi^0\partial^{\mu}\pi^0
-\frac{1}{2}\bar{m}_a^2a^2-\frac{1}{2}\bar{m}_{\pi}^2(1+\delta_{a\pi}^2)\pi^0\pi^0\,.
\end{equation}
The correction term $\delta_{a\pi}^2$ in the pion mass behaves $\mathcal{O}(1/f_a^4)$ and can be safely ignored. Then the LO propagators of $a$ and $\pi$ read 
\begin{equation}
\bar{G}=
\begin{pmatrix}
\bar{G}_{aa}&0\\0&\bar{G}_{\pi^0\pi^0}
\end{pmatrix}\,,
\end{equation}
with
\begin{equation}
\bar{G}_{aa}=\frac{i}{p^2-\bar{m}_a^2+i0^+}\,,
\quad
\bar{G}_{\pi^0\pi^0}=\frac{i}{p^2-\bar{m}_{\pi}^2+i0^+}\,.
\end{equation}
With contributions from the various two-point 1PI amplitudes $\Sigma_{ij}$, the two-point Green functions of $a$ and $\pi$ take the form
\begin{equation}
G=\bar{G}.\left[1-(-i\Sigma).\bar{G}\right]^{-1}\,,
\end{equation}
with
\begin{equation}
G=
\begin{pmatrix}
G_{aa}&G_{a\pi^0}\\G_{a\pi^0}&G_{\pi^0\pi^0}
\end{pmatrix}\,,
\quad
\Sigma=
\begin{pmatrix}
\Sigma_{aa}&\Sigma_{a\pi^0}\\\Sigma_{a\pi^0}&\Sigma_{\pi^0\pi^0}
\end{pmatrix}\,.
\end{equation}
Up to NLO in chiral expansion, the explicit expressions of the matrix elements of $G$ and $\Sigma$ read
\begin{align}
&G_{aa}=\frac{i}{p^2-\bar{m}_a^2-\Sigma_{aa}^{(4)}+i0^+}\,,
\\
&G_{\pi^0\pi^0}=\frac{i}{p^2-\bar{m}_{\pi}^2-\Sigma_{\pi^0\pi^0}^{(4)}+i0^+}\,,
\\
&G_{a\pi^0}=
\frac{i\Sigma_{a\pi^0}^{(4)}}
{(p^2-\bar{m}_a^2-\Sigma_{aa}^{(4)}+i0^+)(p^2-\bar{m}_{\pi}^2-\Sigma_{\pi^0\pi^0}^{(4)}+i0^+)}\,,\label{EQ-B12}
\end{align}
where 
\begin{align}
&\Sigma_{aa}^{(4)}(p^2)=\frac{3\gamma_{ud}}{2f_a^2}\bar{m}_{\pi}^2F_{\beta}(T)
+2(l_3+h_1-h_3)\frac{\gamma_{ud}}{f_a^2}\bar{m}_{\pi}^4-8l_7\frac{\gamma_{ud}^2}{f_a^2}\bar{m}_{\pi}^4\,,
\\
&\Sigma_{\pi^0\pi^0}^{(4)}(p^2)=-\frac{1}{6F^2}(4p^2-\bar{m}_{\pi}^2)F_{\beta}(T)
-\frac{2}{F^2}l_4\bar{m}_{\pi}^2p^2+\frac{2}{F^2}(l_3+l_4)\bar{m}_{\pi}^4-\frac{2\delta_I^2}{F^2}l_7\bar{m}_{\pi}^4\,,
\\
&\Sigma_{a\pi^0}^{(4)}(p^2)=-\delta_I\frac{2}{3Ff_a}p^2 F_{\beta}(T)-\delta_I\frac{1}{Ff_a}l_4\bar{m}_{\pi}^2p^2+\delta_I\frac{4\gamma_{ud}}{Ff_a}l_7\bar{m}_{\pi}^4\,,
\end{align}
and the superscript $(4)$ denotes the $\mathcal{O}(p^4)$ contribution. In this work, the isospin breaking correction in $m_{\pi}$ is neglected, and therefore we have 
\begin{equation}
\begin{aligned}\label{EQ-B16}
&\Sigma_{\pi\pi}^{(4)}(p^2)=\Sigma_{\pi^{\pm}}^{(4)}(p^2)=\Sigma_{\pi^0\pi^0}^{(4)}(p^2)
\\
=&-\frac{1}{6F^2}(4p^2-\bar{m}_{\pi}^2)F_{\beta}(T)
-\frac{2}{F^2}l_4\bar{m}_{\pi}^2p^2+\frac{2}{F^2}(l_3+l_4)\bar{m}_{\pi}^4\,.
\end{aligned}
\end{equation}
In the calculation of the axion-pion scattering amplitudes one will need $G_{a\pi^0}(m_a^2)$ to account for the $a$-$\pi^0$ mixing at NLO. Ignoring the terms of $\mathcal{O}(1/f_a^2)$ and pulling out the axion pole in Eq.~(\ref{EQ-B12}), by keeping the terms up to NLO one has
\begin{equation}
G_{a\pi^0}(p^2)\big|_{p^2\to 0}=\frac{i}{p^2+i0^+}\left(-\frac{\Sigma_{a\pi^0}^{(4)}(0)}{m_{\pi}^2}\right)\,.
\end{equation}

\begin{figure}[t]
	\centering
	\includegraphics[width=0.7\linewidth]{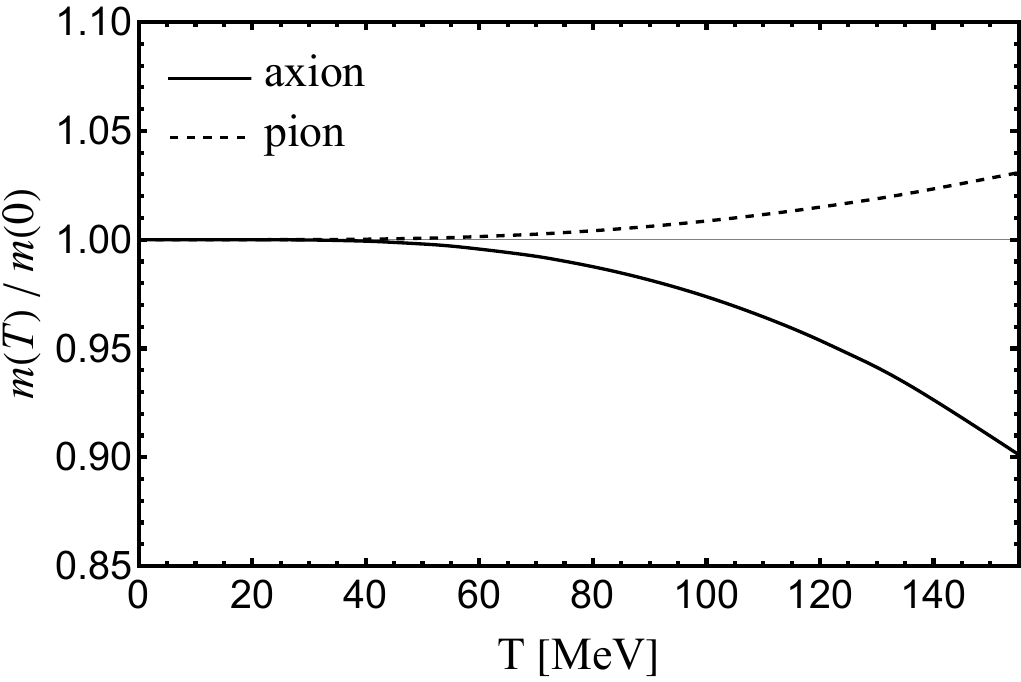}
	\caption{The ratios of the axion and pion masses at finite and zero temperatures predicted by the NLO $\xpt$.}\label{fig.massT}
\end{figure}

Up to NLO in $\xpt$, the ratios of the pion and axion masses at finite and zero temperatures can be written as  
\begin{align}
\frac{m_{\pi}^2(T)}{m_\pi^2(0)}= 1-\frac{1}{2F^2}{\Delta}F_{\beta}(T)\,,
\qquad
\frac{m_a^2(T)}{m_a^2(0)}=1+\frac{3}{2F^2}\Delta F_{\beta}(T)\,,
\end{align}
where the masses at zero temperature read 
\begin{align}
&m_{\pi}^2(0)=\bar{m}_{\pi}^2\left[1-\bar{l}_3\frac{\bar{m}_{\pi}^2}{2(4\pi F)^2}\right]\,,
\\
&m_a^2(0)=\gamma_{ud}\,m_{\pi}^2\frac{F^2}{f_a^2}
\left\{
1-2\frac{m_{\pi}^2}{(4\pi F)^2}\log\frac{m_{\pi}^2}{\mu^2}+2\left[h_1^r(\mu^2)-h_3\right]\frac{m_{\pi}^2}{F^2}
-8l_7\gamma_{ud}\frac{m_{\pi}^2}{F^2}
\right\}\,.\label{eq.maNLO}
\end{align}
In Eq.~(\ref{eq.maNLO}) we have replaced $\bar{m}_{\pi}^2$ with the ``physical'' pion mass $m_{\pi}^2$ up to NLO. The results are consistent with  Ref.~\cite{GrillidiCortona:2015jxo} by ignoring $\delta_I$ term in $m_{\pi}^2$. When calculating the boundary of the axion mass, the renormalization scale $\mu$ is set at 770~MeV, and $h_1=h_3=0$ is taken.  Our result confirms the temperature dempendence of the QCD axion mass in the former reference as well. 
The curves of $m_a(T)/m_a(T=0)$ and $m_{\pi}(T)/m_{\pi}(T=0)$  are given in Fig.~\ref{fig.massT}. 
As shown in this figure, the thermal correction decreases the axion mass around $10\%$ up to $T=155$~MeV, while it slightly increases the pion mass at the level around $3\%$.

\section{Explicit expressions of the $a\pi \to \pi\pi$ scattering amplitudes at finite temperatures }\label{appendix.C}
\setcounter{equation}{0}
\def\theequation{C.\arabic{equation}}

Here we give the explicit one-loop expressions of the $a\pi^0 \to \pi^+\pi^-$ and $a\pi^0 \to \pi^0\pi^0$ scattering amplitudes at finite temperatures. For completeness we also give the zero-temperature parts of the scattering amplitudes, although they have been given in Ref.~\cite{DiLuzio:2022gsc}.  The finite-temperature corrections to the $a\pi \to \pi\pi$ scattering amplitudes represent the new ingredients of this work.

\subsection{$a(p_1)\pi^0(p_2) \to \pi^+(p_3)\pi^-(p_4)$ scattering amplitudes}

The LO amplitude for $a\pi^0\to\pi^+\pi^-$ is
\begin{equation}
\mathcal{M}_{a\pi^0;\pi^+\pi^-}^{(2)}=\frac{\delta_I}{2f_aF}(s-m_{\pi}^2)\,.
\end{equation}
In this work, we will use in the scattering amplitudes the LO pion decay constant $F$, which is independent of the temperature $T$. Up to NLO, $F$ is related to the physical pion decay constant $F_\pi$ via $F=F_{\pi}\left[1-\frac{m_{\pi}^2}{(4\pi F_{\pi})^2}\bar{l}_4\right]$~\cite{Gasser:1983yg}, which leads to $F=86.0$~MeV by taking the same value of $\bar{l}_4=4.73$ from Ref.~\cite{DiLuzio:2022gsc}.

The NLO amplitude is decomposed into the zero-temperature part and finite-temperature correction part. The zero-temperature part is~\cite{DiLuzio:2022gsc}
\begin{align}
\mathcal{M}_{a\pi^0;\pi^+\pi^-}^{(4),\,(T=0)}=&
\frac{\delta_I}{192 \pi^2 f_a F^3} \Big\{
2\bar{l}_1(s-m_{\pi}^2)(s-2m_{\pi}^2)+2\bar{l}_2\left[4m_{\pi}^4+t^2+u^2-3m_{\pi}^2(t+u)\right]\notag
\\
&+\frac{1}{3}\left[11t^2+8tu+11u^2-15m_{\pi}^2(t+u)\right]
+3s(s-m_{\pi}^2)\sigma_{\pi}(s)\log\left(\frac{\sigma_{\pi}(s)-1}{\sigma_{\pi}(s)+1}\right)\notag
\\
&+\left[9m_{\pi}^4-4m_{\pi}^2(s+2t)+t(s+2t)\right]
\sigma_{\pi}(t)\log\left(\frac{\sigma_{\pi}(t)-1}{\sigma_{\pi}(t)+1}\right)\notag
\\
&\left.+\left[9m_{\pi}^4-4m_{\pi}^2(s+2u)+u(s+2u)\right]
\sigma_{\pi}(u)\log\left(\frac{\sigma_{\pi}(u)-1}{\sigma_{\pi}(u)+1}\right)
\right\}\notag
\\
&-\frac{4\gamma_{ud} \delta_I}{f_a F^3}\,l_7m_{\pi}^2(s-2m_{\pi}^2)\,,
\end{align}
with $\sigma_{\pi}(s)=\sqrt{1-4m_{\pi}^2/s}$\,. The finite-temperature correction part is
\begin{equation}
\begin{aligned}\label{eq.ftmap0pppm}
\Delta\mathcal{M}_{a\pi^0;\pi^+\pi^-}^{(4),\,(T)}=&
\frac{3}{2}\Delta\Sigma_{\pi\pi}^{(4)'}(m_{\pi}^2) \mathcal{M}_{a\pi^0;\pi^+\pi^-}^{(2)}
+\Delta A_{a\pi^0;\pi^+\pi^-}^{(4)\text{Tad}}
+\Delta A_{a\pi^0;\pi^+\pi^-}^{(4)s}
\\
&+\Delta A_{a\pi^0;\pi^+\pi^-}^{(4)t}+\Delta A_{a\pi^0;\pi^+\pi^-}^{(4)u}\,,
\end{aligned}
\end{equation}
where ${\Delta}\Sigma_{\pi\pi}^{(4)'}$ is the finite-temperature part of the derivative of Eq.~(\ref{EQ-B16})
\begin{equation}
{\Delta}\Sigma_{\pi\pi}^{(4)'}(m_{\pi}^2)=-\frac{2}{3F^2}{\Delta}F_{\beta}(T)\,.
\end{equation} 
The last four terms of Eq.~\eqref{eq.ftmap0pppm} correspond to the finite-temperature correction parts from the tadpole diagram~\ref{fig.1c}, the $s$-(\ref{fig.1d}), $t$-(~\ref{fig.1e}), and $u$-channel (~\ref{fig.1f}) diagrams, respectively, whose expressions read
\begin{align}
{\Delta}A_{a\pi^0;\pi^+\pi^-}^{(4)\text{Tad}}=&
\frac{5\delta_I}{6f_aF^3}(s-m_{\pi}^2)\Delta F_{\beta}(T)\,,
\\
{\Delta}A_{a\pi^0;\pi^+\pi^-}^{(4)s}=&
\frac{\delta_I}{6f_aF^3}\Big[
s(s-m_{\pi}^2)\Delta J_{0}(T,p_s^0,|\vec{p}_s|)
+p_s^0(s-m_{\pi}^2)\Delta J_{1}(T,p_s^0,|\vec{p}_s|)\notag
\\
&-(s-m_{\pi}^2)\Delta F_{\beta}(T)
-|\vec{p}_s|^2(s-m_{\pi}^2)\Delta b_1(T,p_s^0,|\vec{p}_s|)
\Big]\,,\label{EQ-C6}
\\
{\Delta}A_{a\pi^0;\pi^+\pi^-}^{(4)t}=&\frac{\delta_I}{f_aF^3}
\Big\{
\frac{1}{12}(t-m_{\pi}^2)(t-3m_{\pi}^2)\Delta J_{0}(T,p_t^0,|\vec{p}_t|)
-\frac{1}{12}(t-m_{\pi}^2)\Delta F_{\beta}(T)\notag
\\
&+\frac{1}{6}(t-m_{\pi}^2)(p_2^0+2p_4^0)\Delta J_{1}(T,p_t^0,|\vec{p}_t|)
-p_1^0(p_2^0+p_4^0)\Delta J_{2}(T,p_t^0,|\vec{p}_t|)\notag
\\
&-\frac{1}{6}(t-m_{\pi}^2)\left[(\vec{p}_2+2\vec{p}_4)\cdot \vec{p}_t\right]
\Delta b_1(T,p_t^0,|\vec{p}_t|)\notag
\\
&+\left[p_1^0(\vec{p}_2+\vec{p}_4)\cdot \vec{p}_t+(p_2^0+p_4^0)\vec{p}_1\cdot\vec{p}_t\right]
\Delta b_{J1}(T,p_t^0,|\vec{p}_t|)\notag
\\
&-\vec{p}_1\cdot(\vec{p}_2+\vec{p}_4)\Delta b_{21}(T,p_t^0,|\vec{p}_t|)\notag
\\
&-\vec{p}_1\cdot\vec{p}_t\left[(\vec{p}_2+\vec{p}_4)\cdot\vec{p}_t\right]
\Delta b_{20}(T,p_t^0,|\vec{p}_t|)
\Big\}\,,
\\
{\Delta}A_{a\pi^0;\pi^+\pi^-}^{(4)u}=&\frac{\delta_I}{f_aF^3}
\Big\{
\frac{1}{12}(u-m_{\pi}^2)(u-3m_{\pi}^2)\Delta J_{0}(T,p_u^0,|\vec{p}_u|)
-\frac{1}{12}(u-m_{\pi}^2)\Delta F_{\beta}(T)\notag
\\
&+\frac{1}{6}(u-m_{\pi}^2)(p_2^0+2p_3^0)\Delta J_{1}(T,p_u^0,|\vec{p}_u|)
-p_1^0(p_2^0+p_3^0)\Delta J_{2}(T,p_u^0,|\vec{p}_u|)\notag
\\
&-\frac{1}{6}(u-m_{\pi}^2)\left[(\vec{p}_2+2\vec{p}_3)\cdot \vec{p}_u\right]
\Delta b_1(T,p_u^0,|\vec{p}_u|)\notag
\\
&+\left[p_1^0(\vec{p}_2+\vec{p}_3)\cdot\vec{p}_u+(p_2^0+p_3^0)\vec{p}_1\cdot\vec{p}_u\right]
\Delta b_{J1}(T,p_u^0,|\vec{p}_u|)\notag
\\
&-\vec{p}_1\cdot(\vec{p}_2+\vec{p}_3)\Delta b_{21}(T,p_u^0,|\vec{p}_u|)\notag
\\
&-\vec{p}_1\cdot\vec{p}_u\left[(\vec{p}_2+\vec{p}_3)\cdot\vec{p}_u\right]
\Delta b_{20}(T,p_u^0,|\vec{p}_u|)
\Big\}\,.\label{EQ-C8}
\end{align}

\subsection{$a(p_1)\pi^0(p_2) \to \pi^0(p_3)\pi^0(p_4)$ scattering amplitudes}

The nonvanishing contribution to the $a\pi^0 \to \pi^0\pi^0$ scattering amplitude begins at $\mathcal{O}(p^4)$. The zero temperature part is
\begin{equation}
\begin{aligned}
\mathcal{M}_{a\pi^0;\pi^0\pi^0}^{(4),\,(T=0)}=&\frac{\delta_I}{32\pi^2 f_a F^3} 
\Big[
\frac{1}{3}(\bar{l}_1+2\bar{l}_2)(s^2+t^2+u^2-3m_{\pi}^4)+(s^2+t^2+u^2-3m_{\pi}^4)\label{eqA8}
\\
&+(s-m_{\pi}^2)^2\sigma_{\pi}(s)\log\left(\frac{\sigma_{\pi}(s)-1}{\sigma_{\pi}(s)+1}\right)
+(t-m_{\pi}^2)^2\sigma_{\pi}(t)\log\left(\frac{\sigma_{\pi}(t)-1}{\sigma_{\pi}(t)+1}\right)
\\
&+(u-m_{\pi}^2)^2\sigma_{\pi}(u)\log\left(\frac{\sigma_{\pi}(u)-1}{\sigma_{\pi}(u)+1}\right)
\Big]
+\frac{12\gamma_{ud}\delta_I}{f_a F^3} l_7 m_{\pi}^4\,.
\end{aligned}
\end{equation}
It is noted that there is a typo for the $l_7$ term in Ref.~\cite{DiLuzio:2022gsc} and the numeric coefficient should be 12,  instead of 36.~\footnote{We thank Gioacchino Piazza for confirming this coefficient.} 
The finite-temperature correction part of the amplitude is
\begin{equation}\label{EQ-C10}
\begin{aligned}
\Delta\mathcal{M}_{a\pi^0;\pi^0\pi^0}^{(4),\,(T)}=&\frac{\delta_I}{2f_aF^3}
\Big[
(s-m_{\pi}^2)^2\Delta J_{0}(T,p_s^0,|\vec{p}_s|)
+(t-m_{\pi}^2)^2\Delta J_{0}(T,p_t^0,|\vec{p}_t|)\\
&+(u-m_{\pi}^2)^2\Delta J_{0}(T,p_u^0,|\vec{p}_u|)
\Big]\,.
\end{aligned}
\end{equation}

\section{Phase space integrals in the axion thermalization rate}\label{appendix.D}
\setcounter{equation}{0}
\def\theequation{D.\arabic{equation}}

\begin{figure*}[htp]
	\centering
	\includegraphics[width=0.3\linewidth]{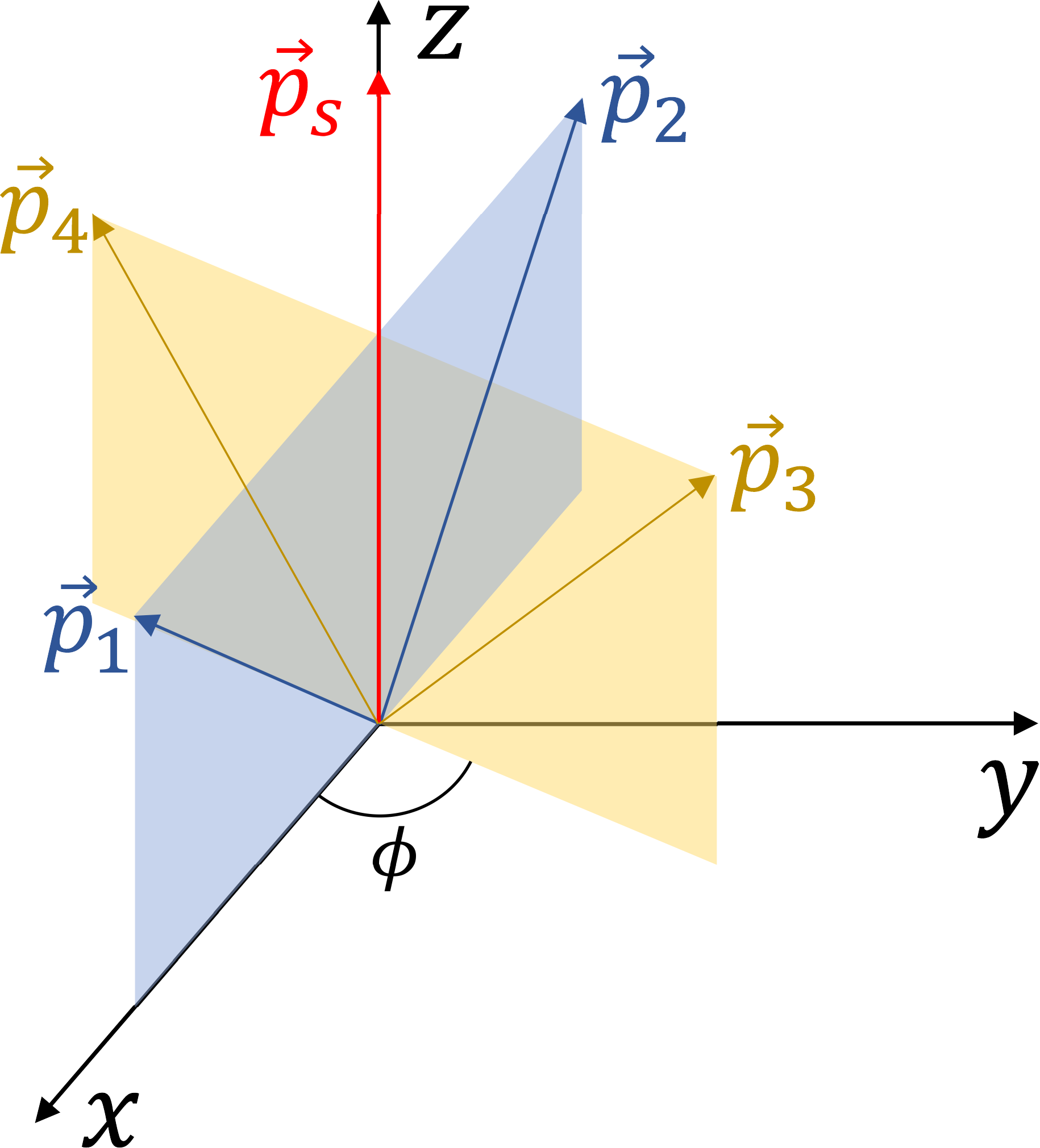}
	\caption{Coordinate system chosen for the evaluation of the phase space integral.}\label{fig.frame}
\end{figure*}

We follow the steps of Ref.~\cite{Green:2021hjh} to reduce the twelve-dimensional phase space integral in the axion thermalization rate to a five-dimensional one. It is convenient to use $p_s=p_1+p_2=(E_s,\,\vec{p}_s)$ for illustration. Due to the rotational invariance, the integrand of Eq.~\eqref{eq.gammat} does not depend on the direction of $\vec{p}_s$ and thus we can take the $\vec{z}$ direction as $\hat{\vec{z}}=\vec{p}_s/|\vec{p}_s|$, as shown in Fig.~\ref{fig.frame}. Without loss of generality, one can assign the $(\vec{p}_1,\,\vec{p}_2)$ plane as the $(\hat{\vec{x}},\,\hat{\vec{z}}\,)$ plane. And $\phi$ stands for the angle between the $(\vec{p}_3,\,\vec{p}_4)$ plane and the $(\vec{p}_1,\,\vec{p}_2)$ plane.

By defining $\vec{p}_i\cdot\vec{p}_s=|\vec{p}_i|\,|\vec{p}_s|\cos\theta_i$ and using the three-momenta conservation, one has
\begin{equation}
\begin{aligned}
&\cos\theta_1=\frac{|\vec{p}_s|^2+|\vec{p}_1|^2-|\vec{p}_2|^2}{2|\vec{p}_s||\vec{p}_1|}\,,
\quad
\cos\theta_2=\frac{|\vec{p}_s|^2+|\vec{p}_2|^2-|\vec{p}_1|^2}{2|\vec{p}_s||\vec{p}_2|}\,,
\\
&\cos\theta_3=\frac{|\vec{p}_s|^2+|\vec{p}_3|^2-|\vec{p}_4|^2}{2|\vec{p}_s||\vec{p}_3|}\,,
\quad
\cos\theta_4=\frac{|\vec{p}_s|^2+|\vec{p}_4|^2-|\vec{p}_3|^2}{2|\vec{p}_s||\vec{p}_4|}\,.
\end{aligned}
\end{equation} 
According to the coordinate of Fig.~\ref{fig.frame}, each component of $\vec{p}_i$ can be written as  
\begin{align}
&\vec{p}_1=|\vec{p}_1|(\sin\theta_1,\,0,\,\cos\theta_1)\,,\label{EQ-D3}
\\
&\vec{p}_2=|\vec{p}_2|(-\sin\theta_2,\,0,\,\cos\theta_2)\,,
\\
&\vec{p}_3=|\vec{p}_3|(\sin\theta_3\cos\phi,\,\sin\theta_3\sin\phi,\,\cos\theta_3)\,,
\\
&\vec{p}_4=|\vec{p}_4|(-\sin\theta_4\cos\phi,\,-\sin\theta_4\sin\phi,\,\cos\theta_4)\,.\label{EQ-D6}
\end{align}

Next we reduce the twelve-dimensional integral of Eq.~(\ref{eq.PSInt}) to a five-dimensional one. In the first step we substitute the following identity
\begin{equation}
1=\int_0^{+\infty}\mathrm{d}E_s\int\mathrm{d}^3p_s\delta(E_s-E_1-E_2)\delta^3(\vec{p}_s-\vec{p}_1-\vec{p}_2)\,,
\end{equation}
into the Eq.~(\ref{eq.PSInt}), and then $\vec{p}_2$ and $\vec{p}_4$ can be integrated out by the two $\delta$ functions of three-momenta conservation
\begin{align}
\int\mathrm{d}\widetilde{\Gamma}=&
\int_0^{+\infty}\mathrm{d}E_s\int\mathrm{d}^3p_s\int\mathrm{d}^3p_1\int\mathrm{d}^3p_3
\frac{\prod_{i=1,3}\delta(E_s-E_i-E_{i+1})}{16(2\pi)^8E_1E_2E_3E_4}\bigg|_{\vec{p}_2=\vec{p}_s-\vec{p}_1,\,\vec{p}_4=\vec{p}_s-\vec{p}_3}
\\
=&\int_0^{+\infty}\mathrm{d}E_s\int_0^{+\infty}\mathrm{d}|\vec{p}_s|\int_0^{+\infty}\mathrm{d}|\vec{p}_1|\int_0^{+\infty}\mathrm{d}|\vec{p}_3|\int_0^{2\pi}\mathrm{d}\phi\left(\int_{-1}^{+1}\prod_{i=1,3}\mathrm{d}\cos\theta_i\right)\notag
\\
&\times\left[\prod_{i=1,3}\delta(E_s-E_i-E_{i+1})\right]
\frac{|\vec{p}_s|^2|\vec{p}_1|^2|\vec{p}_3|^2}{8(2\pi)^6E_1E_2E_3E_4}\,.\notag
\end{align}
The integrals over $\cos\theta_{1,3}$ can be performed by the remaining two $\delta$ functions. Using $\delta\left(g(x)\right)=\sum_i\delta(x-\bar{x}_i)/|g'(\bar{x}_i)|$ with $\bar{x}_i$ the root of $g(x)$\,, we have
\begin{equation}
\prod_{i=1,3}\delta(E_s-E_i-E_{i+1})
=\frac{E_2E_4}{|\vec{p}_s|^2|\vec{p}_1||\vec{p}_3|}
\prod_{i=1,3}\delta\left(\cos\theta_i-\frac{2E_s\sqrt{|\vec{p}_i|^2+m_i^2}-(s+m_i^2-m_{i+1}^2)}{2|\vec{p}_s||\vec{p}_i|}\right)\,,
\end{equation}
with $s=E_s^2-|\vec{p}_s|^2$\,, which requires that
\begin{equation}
\left|\frac{2E_s\sqrt{|\vec{p}_i|^2+m_i^2}-(s+m_i^2-m_{i+1}^2)}{2|\vec{p}_s||\vec{p}_i|}\right|\leq1\,.
\end{equation}
Expanding the above inequation one can get
\begin{equation}
16s^2|\vec{p}_i|^4+(8sB_i-16A_i^2E_s^2)|\vec{p}_i|^2+B_i^2-16A_i^2E_s^2m_i^2\leq0\,,\label{EQ-D11}
\end{equation}
with $A_i=s+m_i^2-m_{i+1}^2$ and $B_i=A_i^2+4E_s^2m_i^2$\,. The existence of solutions for the quadratic inequality about $|\vec{p}_i|^2$ requires $s\geq(m_i+m_{i+1})^2$, which is equivalent to 
\begin{equation}
E_s\geq{m_i+m_{i+1}}\,,\quad|\vec{p}_s|\leq\sqrt{E_s^2-(m_i+m_{i+1})^2}\,.
\end{equation}
For the cases of $i=1$ and $3$\,, the requirements for $E_s$ and $|\vec{p}_s|$ are
\begin{equation}\label{EQ-D13}
E_s^{\text{min}}=\max_{i=1,3}\left\{m_i+m_{i+1}\right\}\,,
\quad
|\vec{p}_s|_{\text{max}}=\min_{i=1,3}\left\{
\sqrt{E_s^2-(m_i+m_{i+1})^2}
\right\}\,.
\end{equation}
The solutions of Eq.~(\ref{EQ-D11}) provide the integral boundaries to the $|\vec{p}_{i=1,3}|$ variables~\cite{Green:2021hjh}
\begin{equation}\label{EQ-D14}
|\vec{p}_i|_{\text{min,max}}=\frac{1}{2s}\left|
E_s\sqrt{\left[s-(m_i-m_{i+1})^2\right]\left[s-(m_i+m_{i+1})^2\right]}\mp(s+m_i^2-m_{i+1}^2)|\vec{p}_s|
\right|\,.
\end{equation}
Therefore the phase space integral in the axion thermalization rate can be now written as a five-dimensional integral
\begin{equation}\label{EQ-D15}
\int\mathrm{d}\widetilde{\Gamma}=
\int_{E_s^{\text{min}}}^{+\infty}\mathrm{d}E_s\int^{|\vec{p}_s|_{\text{max}}}_0\mathrm{d}|\vec{p}_s|
\int_{|\vec{p}_1|_{\text{min}}}^{|\vec{p}_1|_{\text{max}}}\mathrm{d}|\vec{p}_1|
\int_{|\vec{p}_3|_{\text{min}}}^{|\vec{p}_3|_{\text{max}}}\mathrm{d}|\vec{p}_3|
\int_0^{2\pi}\mathrm{d}\phi\,\frac{|\vec{p}_1||\vec{p}_3|}{8(2\pi)^6E_1E_3}\,,
\end{equation}
with the integral boundaries given in Eqs.~(\ref{EQ-D13}) and (\ref{EQ-D14}).
All the kinematic variables in thermal scattering amplitudes given in Appendix~\ref{appendix.C} can be expressed by the five integral variables in Eq.~(\ref{EQ-D15}) through Eqs.~(\ref{EQ-D3})-(\ref{EQ-D6}).

\bibliography{axion-pion-scattering-arxiv}
\bibliographystyle{apsrev4-2}

\end{document}